\documentclass[prd,aps,floatf,twocolumn,showpacs,color]{revtex4}
\usepackage{amsmath}
\usepackage{amssymb}
\usepackage{mathrsfs}
\usepackage{natbib}
\usepackage{graphicx}
\newcommand\solidrule[1][0.7cm]{\rule[0.5ex]{#1}{.4pt}}
\newcommand\dashedrule{\mbox{\solidrule[2mm]\hspace{1mm}\solidrule[2mm]\hspace{1mm}\solidrule[2mm]}}
\newcommand\dasheddotrule{\mbox{\solidrule[2mm]\hspace{1mm}$\cdot$\hspace{1mm}\solidrule[2mm]}}
\begin{document}
\title{\large \bf Gravitational Lensing by a structure in a cosmological background}
\author{M. Parsi Mood}
\affiliation{Department of Physics, Sharif University of Technology,
Tehran, Iran} \email{parsimood@physics.sharif.edu}

\author{Javad T. Firouzjaee}
\affiliation{School of Astronomy and Physics, Institute for Research in Fundamental Sciences (IPM), Tehran, Iran,
Tehran, Iran}
\author{Reza Mansouri}
\affiliation{Department of Physics, Sharif University of Technology,
Tehran, Iran and \\
  School of Astronomy, Institute for Research in Fundamental Sciences (IPM), Tehran, Iran}
\email{mansouri@ipm.ir}

\date{\today}

\begin{abstract}
We use an exact general relativistic model structure within an FRW cosmological background based on a LTB metric to study the gravitational lensing by a cosmological and dynamical structure. Using different density profiles for the model structure, the deviation angle and the time delay through the gravitational lensing has been studied by solving the geodesic equations. The results of these exact calculations have been compared to the thin lens approximation. We have shown that the result for the thin lens approximation based on a modified NFW density profile with a void before going over to the FRW background matches very well with the exact general relativistic calculations. However, the thin lens approximation based on a normal NFW profile does differ from the exact relativistic calculation. The difference is more the less compact the structure is. We have also looked at the impact of our calculation on the observational interpretation of arcs in the case of strong lensing and also the reduced shear in the case of weak lensing. No significant difference has been seen in the data available.

\end{abstract}
\pacs{98.80.Jk, 98.62.Js, 98.62.Ck , 95.35.+d}
\maketitle
\section{introduction}
\par
Nowadays gravitational lensing (GL) has been turned out to be a powerful tool in astrophysics and cosmology, from the exoplanet detection to the dark matter, dark energy and Hubble parameter measurements. The strong and weak lensing technology is now widely used and its thin lens (Th-L) approximation procedure is the prevailing method of estimating the mass of large scale structures like galaxies and cluster of galaxies leading to dark matter and dark energy contents of the universe \cite{GLenses}. The current view is that this Th-L approximation, based on fundamental features of light rays in general relativity, is accurate enough at the cosmological scales where we are faced with very weak gravitational fields and potentials. There has been already some attempts to compare the Th-L method to the more exact integration of null geodesics in a perturbed cosmological background (\cite{Sas93,Fut95}, see also \cite{FriKling11} and the references there). Even in the Th-L methode it is assumed that the effect of inhomogeneities only affect angular distances through Swiss-cheese models \cite{GLenses,DR73,Marra:2011ct} and have not important effect on the lens scale. All these attempts lack a full general relativistic model to be compared with the Th-L technology and to conclude about its accuracy.\\

Perturbative techniques may be used in cases where local features are discussed in the presence of the weak gravity\cite{GreenWald}. However, when non-local or quasi-local phenomena are in play we may not be allowed to use the same perturbative method. This is the main question of this paper. Quasi-local concepts, like the mass of a cosmic  structure in contrast to the mass of a test particle, and the concept of the event horizon or boundary of a black hole, are usually sensitive to global assumptions like being asymptotically flat or asymptotically FRW background, irrespective of how weak or strong the gravity may be. For example, the unique exact event horizon of a Schwarzschild black hole is split into an apparent and an approximative event horizon once it is considered as a dynamical cosmic black hole \cite{taghizadeh}. These counter-intuitive effects not detected in the perturbational approach are inherent in the presence of the quasi-local phenomena. The definition of quasi-local mass of a structure in general relativity is another example \cite{Szabados}: In general there are many definitions for it, none of them being preferred. It has already been shown numerically how different various quasi-local mass definitions of a general relativistic structure may be \cite{taghizadeh}. The matching behavior of a spherically symmetric cosmological structure glued to a FRW background with the necessity of a void reflected in its density profile \cite{khakshournia} is another example of a quasi-local phenomenon relevant to the gravitational lensing problem we are going to study in this paper. Quasi-local field irregularities, such as the wall of large voids, may also lead to new relativistic effects not inferred from the familiar approximative approaches, as shown in \cite{Bolejko12}.\\
\par
Our aim in this paper is to study the exact general relativistic gravitational lensing by a spherically symmetric overdensity structure within an FRW universe and compare it with the corresponding Th-L approach to understand more exactly the accuracy of this technology as a perturbation to the exact general relativistic model calculation. Note that Th-L approximation is the basic underlying model used in cosmology to interpret observational data related to the gravitational lensing. Our cosmic structure is defined by an exact solution of the Einstein equations not produced by a cut-and-paste technology to be sure that all dynamical quasi-local effects are taken into account and to avoid the case of freezing of probable effects or features due to the gluing of different manifolds. There is already an exact general relativistic model structure within an FRW universe based on a Lema\^{\i}tre, Tolman and Bondi (LTB) metric \cite{Lem97,Tol34,Bon47} representing an inhomogeneous cosmological model with a source at its center \cite{taghizadeh}. Choosing such a model for an extended lens, we study the gravitational lensing in the dynamical cosmological background in the framework of general relativity. The null geodesic equations of this exact model are integrated numerically to obtain the deflection angle and time delay for different density profiles. The results are then compared to the corresponding thin lens approximation. \\

Section \ref{LTBmetric} is an introduction to the LTB metric and the corresponding geodesic equations. We then go over in section \ref{structmodel} to define a toy model and a modified NFW structure model consistent with a smooth matching to our FRW background. Section \ref{GLLTB} includes definitions we need to calculate the deflection angle and the time delay either by integrating the geodesic equations or by the corresponding thin lens approximation. The results are then given in section \ref{res}. We then conclude in section \ref{concl}.

\section{LTB metric}\label{LTBmetric}

The metric of an inhomogeneous spherically symmetric LTB space-time may be written in the comoving coordinates as ($G=1,c=1$)
\begin{equation}\label{ltbm}
 ds^{2}=-dt^{2}+X^2(r,t)dr^{2}+R^2(t,r)d\Omega^{2}.
\end{equation}
It represents a cosmological solution filled with pressure-less perfect fluid with density of $\rho(r,t)$ in the absence of a
cosmological constant satisfying
\begin{eqnarray}
\rho(r,t)&=&\frac{M'(r)}{4\pi R^{2}R'},\\
X&=&\frac{R'}{\sqrt{1+E(r)}},\\
\dot{R}^{2}&=&E(r)+\frac{2M(r)}{R}.\label{fieldeqn}
\end{eqnarray}
Here dot and prime denote partial derivatives with respect to the parameters $t$ and $r$ respectively. Equation (\ref{fieldeqn}) has three different analytic solution, depending on the value of $E$, given by
\begin{eqnarray}\label{ltbc}
R&=&-\frac{M}{E}(1-\cos\eta),\nonumber\\
\eta-\sin\eta&=&\frac{(-E)^{3/2}}{M}(t-t_{b}(r)),
\end{eqnarray}
for $E < 0$, and
\begin{equation}\label{ltbef}
R=\left(\frac{9}{2}M(t-t_{b})^2\right)^{\frac{1}{3}},
\end{equation}
 for $E = 0$, and
\begin{eqnarray} \label{ltbeo}
R&=&\frac{M}{E}(\cosh\eta-1),\nonumber\\
\sinh\eta-\eta&=&\frac{E^{3/2}}{M}(t-t_{b}(r)),
\end{eqnarray}
for $E > 0$.
\par
These solutions have three free parameters: $t_{b}(r)$, $E(r)$, and $M(r)$. Given that the metric is covariant under the rescaling $r\rightarrow\tilde{r}(r)$ one of these functions may be fixed.\\
The geodesic equations may be written in the arbitrary plane of $\theta=\frac{\pi}{2}$ due to the spherical symmetry:
\begin{flalign}\label{ltbgeo}
&t:\frac{d^2t}{d\lambda^2} + X\dot{X}\left(\frac{dr}{d\lambda}\right)^2 + R\dot{R}\left(\frac{d\phi}{d\lambda}\right)^2 = 0, \\
&r: \frac{d^2r}{d\lambda^2} + 2\frac{\dot{X}}{X}\frac{dr}{d\lambda}\frac{dt}{d\lambda} + \frac{X'}{X}\left(\frac{dr}{d\lambda}\right)^2-\frac{RR'}{X^2}\left(\frac{d\phi}{d\lambda}\right)^2=0, \\
&\phi: \frac{d^2\phi}{d\lambda^2} + 2\frac{\dot{R}}{R}\frac{dt}{d\lambda}\frac{d\phi}{d\lambda} + 2\frac{R'}{R}\frac{dr}{d\lambda}\frac{d\phi}{d\lambda}=0, \label{geophi}
\end{flalign}
where $\lambda$ is an affine parameter. Equation (\ref{geophi}) expresses the conservation of the angular momentum:
\begin{equation}
L=R^2\frac{d\phi}{d\lambda}=Const.
\end{equation}
We are interested in the light-like geodesics. From the metric we obtain the light-like condition in the form
\begin{equation}\label{null}
\left(\frac{dt}{d\lambda}\right)^2=X^2 \left(\frac{dr}{d\lambda}\right)^2 + R^2 \left(\frac{d\phi}{d\lambda}\right)^2
\end{equation}
 These non-linear coupled differential equations can not be solved analytically. To solve them for specific functions $M(r), t(r)$, and $E(r)$ we do need all derivatives of the metric functions given in the Appendix I.

\section{Structure Modeling}\label{structmodel}

Our study begins with a try to build models of mass condensation within an otherwise expanding universe using the LTB metric and the corresponding field equations. We will start with a toy model expressible in analytic functions and then continue to construct a more realistic model using the procedure proposed in \cite{KH01,BKCH}.

\subsection{Toy model}\label{toymodel}

\begin{figure}[ht]
\includegraphics[width = \columnwidth]{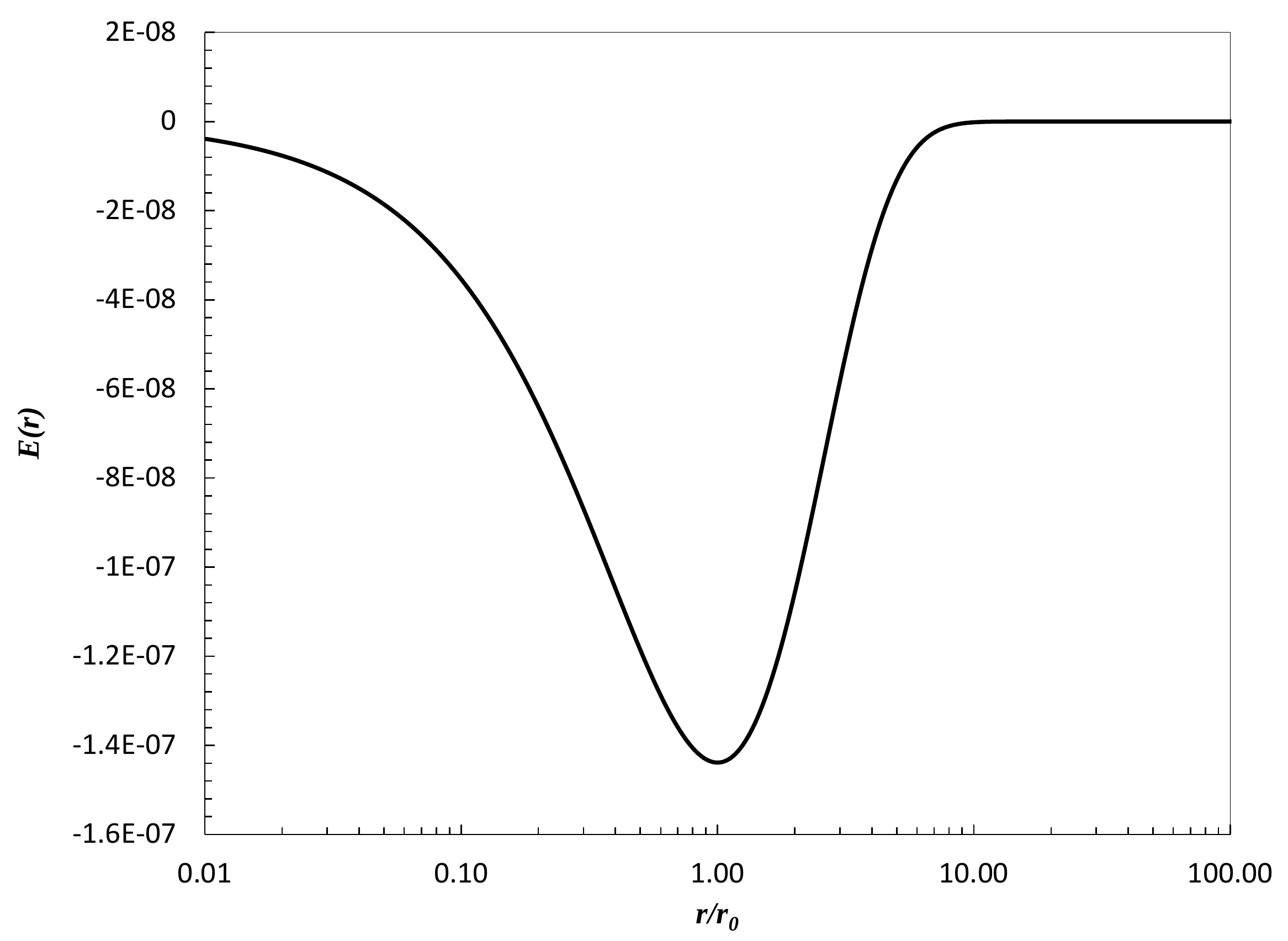}
\hspace*{10mm}\caption{\label{toyer}$E(r)$ for toy model as a function of $\frac{r}{r_0}$.}
\end{figure}

Let's specify some smooth and analytic functions for LTB parameters tending at far distances to the corresponding flat FRW metric and, at the same time, leading to a dense structure at the center. Our choice is a modification of the model proposed in \cite{taghizadeh}:
\begin{eqnarray}
t_b &=& 0 \\
M(r)&=& \frac{1}{a}\left[\left(\frac{r}{r_0}\right)^\frac{3}{2}+ \left(\frac{r}{r_0}\right)^3\right]\\
E(r) &=& -b \left(\frac{r}{r_0}\right)e^{-\frac{r}{r_0}},
\end{eqnarray}
where $a$,$b$ and $r_0$ are positive constants that determine the length scale of structure and its time evolution. The model is mainly characterized by $E$ (see Fig. \ref{toyer}), being negative, having a minimum at $r=r_0$, and approaching zero for large $r$. This ensures us a collapsing structure with a collapse time increasing exponentially with the radius:
\begin{equation}\label{coltime}
t_{col}(r)=\frac{2\pi}{ab}\left[1+\left(\frac{r}{r_0}\right)^\frac{3}{2}\right]e^{\frac{3}{2}\frac{r}{r_0}}.
\end{equation}

Note that for $r=0$ we have $t_{sing}\equiv\frac{2\pi}{ab}$.\\
\par
\begin{figure}[ht]
\includegraphics[width = \columnwidth]{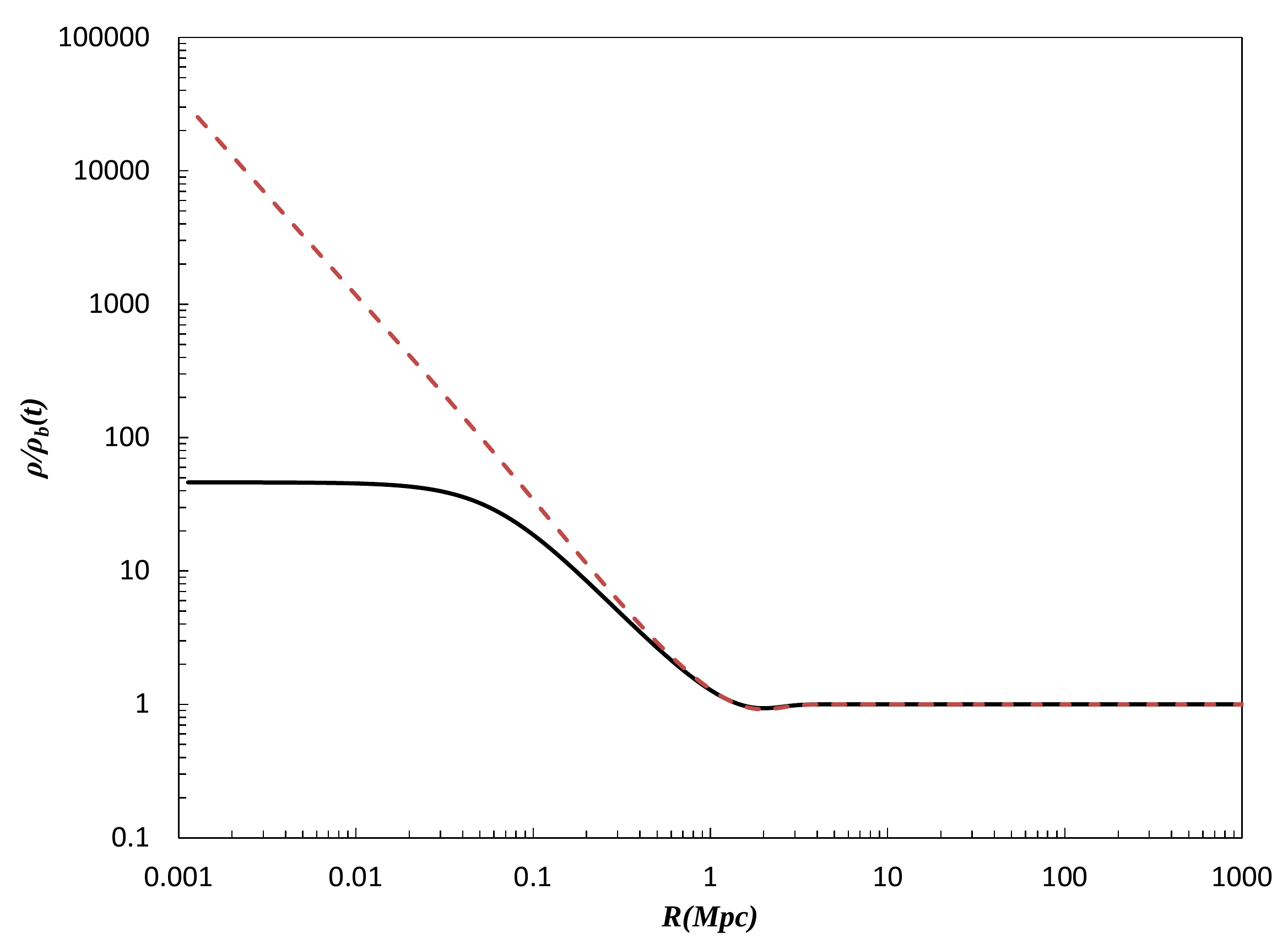}
\hspace*{10mm}\caption{\label{toydenar}Density profile of toy model as a function physical radius $R$ in unit of background density at each time. Solid line which is for some time before $t_{sing}$ is similar to top hat profile and dash line which is for some time after $t_{sing}$ is singular and behave approximately as $R^{-\frac{3}{2}}$ near center. Both profiles tend to background density at far distances.}
\end{figure}

The density profile as a function of the physical radius $R$ at two different times is given in Fig. \ref{toydenar}. It is seen that the density profile of the structure at a time before $t_{sing}$ is very similar to a top hat model. At any time later, however, it becomes singular at the center with a mass in-fall increasing with the time. The profile is then proportional to $R^{-\alpha}$ with $\alpha$ being almost $\frac{3}{2}$. The density at far distances tends to the matter dominated flat FRW background density. We have chosen $a,b$ and $r_0$ such that the scale of the structure is in the range of a cluster of galaxy:
\begin{equation}\label{frwden}
\rho_b(t)=\frac{1}{6\pi t^2}
\end{equation}
The resulting density profile shown in Fig. \ref{toydenar} has a void which is generic to any structure formation in general relativity. We know already that within general relativity it is not possible to paste a spherically symmetric over-dense region to a FRW background except via an under-dense region or a void \cite{khakshournia}. In a more realistic dynamical setting, like a LTB metric as an exact overall solution of Einstein equations we are considering, the numerical calculation shows that any initial density profile representing an over-dense region leads finally to a void before going to the background FRW. This is a general relativistic quasi-local feature which may be related to the voids we know from observations. We have to take it anyhow into consideration in comparing exact general relativistic calculations with any approximative method like the thin lens approximation in the gravitational lensing. Based on this exact result, we will always include a void in our definitions of the density profiles.
\par

\subsection{Model based on a given density profile}\label{NFW model}

\begin{figure}[ht]
\includegraphics[width = \columnwidth]{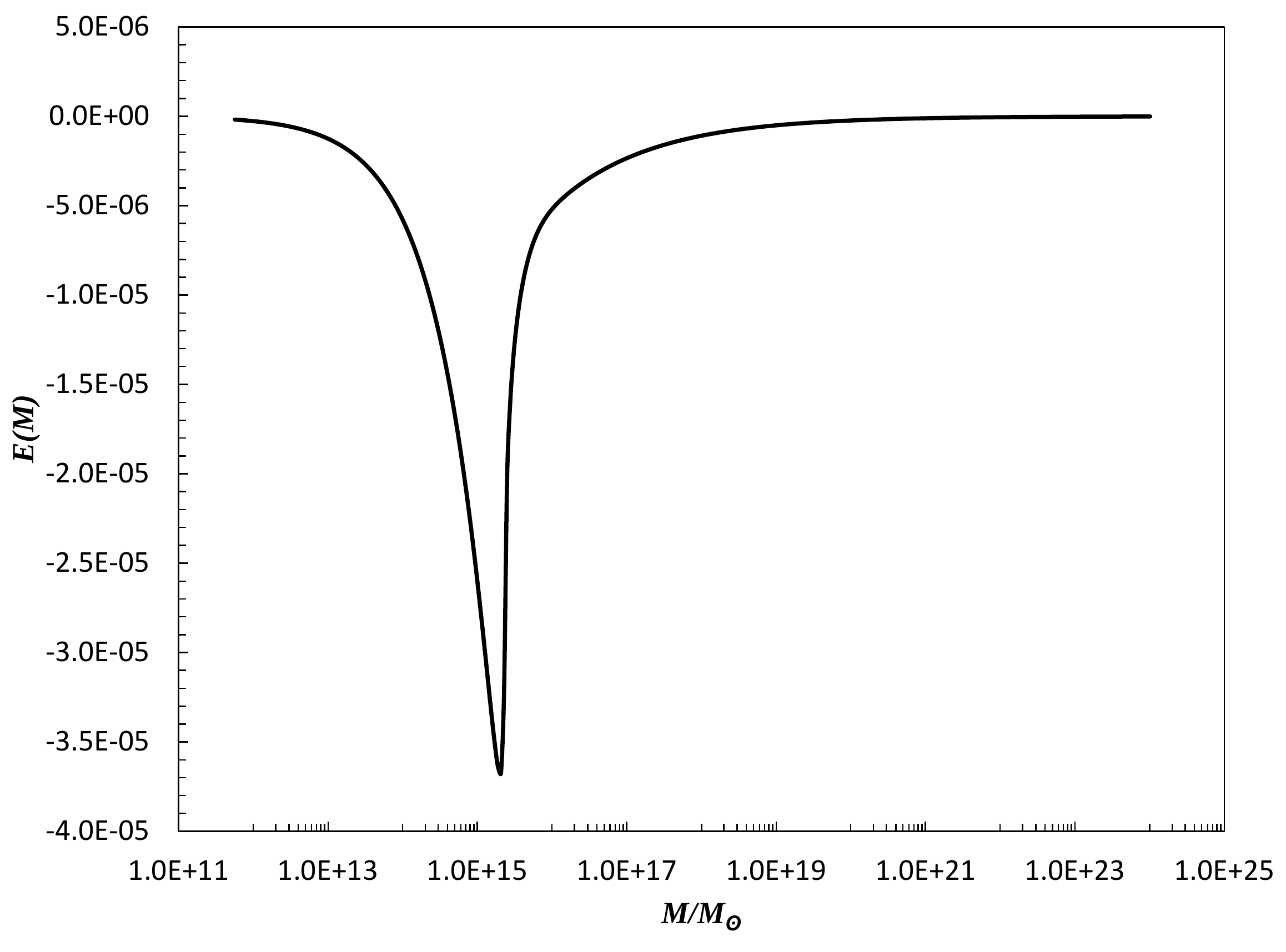}
\hspace*{10mm}\caption{\label{emclnfw}$E$ as a function of $M$ for a cluster with NFW density profile. Unit of $M$ is Sun mass.}
\end{figure}

\begin{figure}[ht]
\includegraphics[width = \columnwidth]{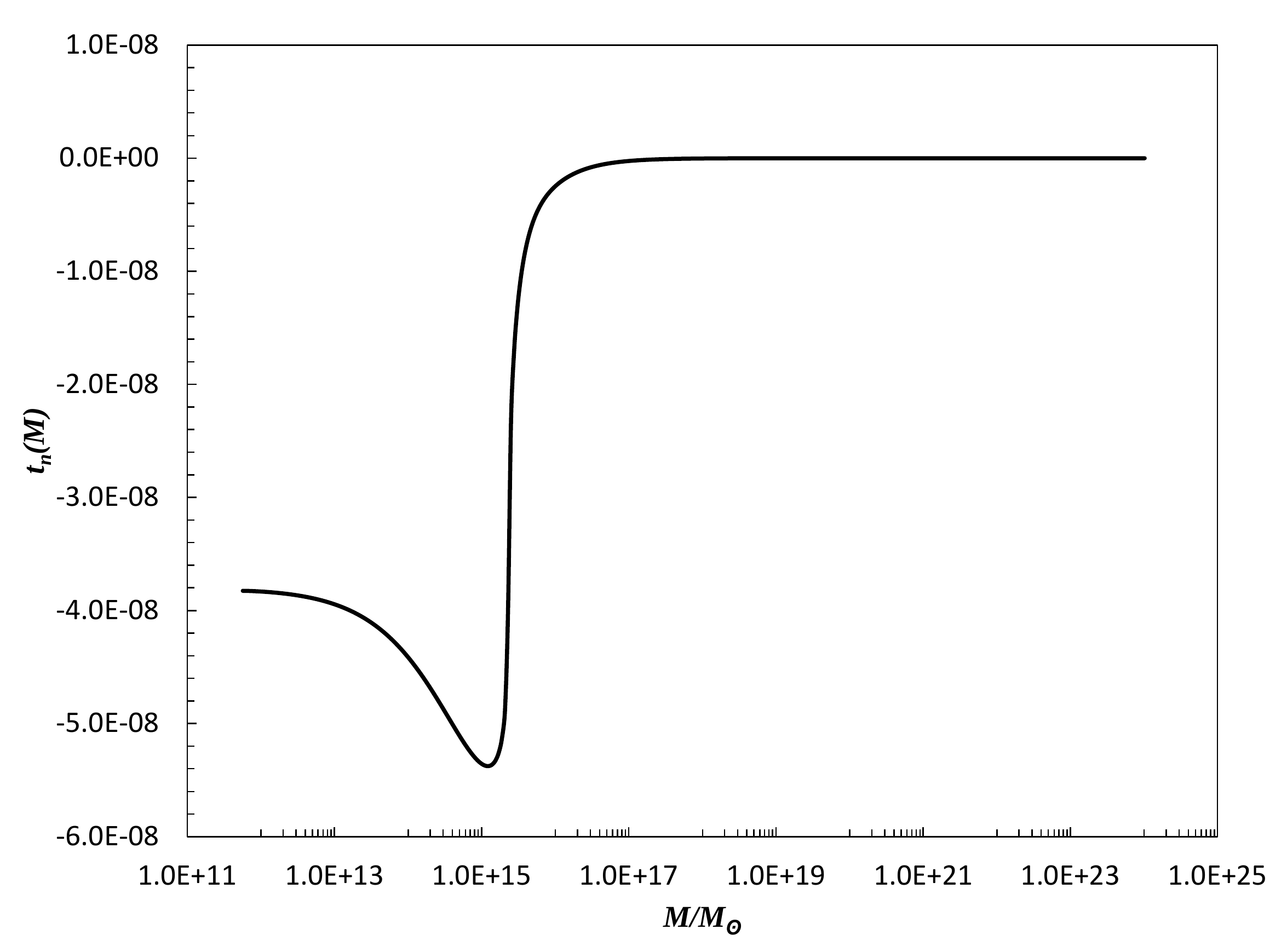}
\hspace*{10mm}\caption{\label{tnclnfw}$t_n$ as a function of $M$ for a cluster with NFW density profile. Unit of $M$ is Sun mass and unit of $t_n$ is $3.263 Gyr$.}
\end{figure}

\begin{figure}[ht]
\includegraphics[width = \columnwidth]{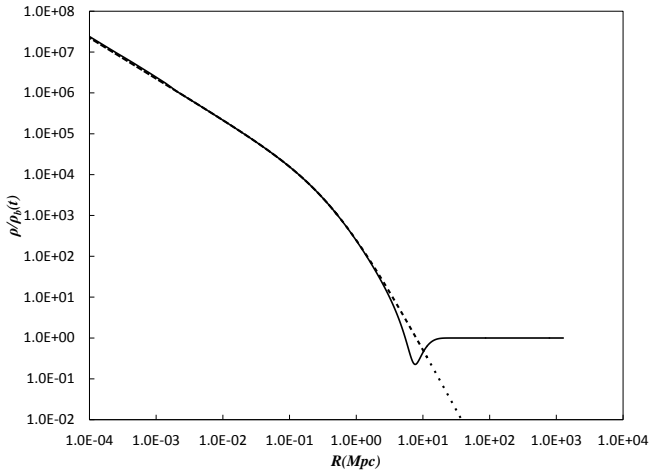}
\hspace*{10mm}\caption{\label{nfwden}Density profile for a cluster. Dot line is for original NFW profile and solid line is for our LTB model (NFW near center that tends to background density via a compensated underdensity).}
\end{figure}

In the case some boundary conditions are given the field equations can be integrated to specify the model. The corresponding algorithm is formulated by Krasi\'{n}ski and Hellaby \cite{KH01,BKCH}. We are interested in a model based on a known density profile. It is then necessary to specify the density profile at two different times $t_1, t_2$ as a function of the coordinate $r$. Now, the algorithm is based on the choice of $r$-coordinate such that $M(r) = r$. This is due to the fact that $M(r)$ is an increasing function of $r$. Therefore, $E$ and $t_n$ become functions of $M$. The LTB functions $E(M)$ and $t_n(M)$ may then be extracted from the algorithm. For the initial time we choose the time of the last scattering surface: $t_1\simeq 3.77\times10^5 yr$. The initial density profile should show a small over-density near the center imitating otherwise a FRW universe. Therefore, we add a Gaussian peak to the FRW background density. We know already that having an over-density in an otherwise homogeneous universe needs a void to compensate for the extra mass within the over-density region. Therefore, to compensate this mass we subtract a wider gaussian peak:
\begin{equation}\label{initden}
\rho(R,t_1)=\rho_b(t_1)\left[\left(\delta_1e^{-\left(\frac{R}{R_0}\right)^2}- b_1\right) e^{-\left(\frac{R}{R_1}\right)^2}+1\right],
\end{equation}
where $\delta_1$ is the density contrast of the Gaussian peak, $R_0$ is the width of the Gaussian peak and $R_1$ is the width of the negative Gaussian profile. The mass compensation condition leads to an equation for $b_1$. For the final time we choose the time when our null geodesy has the nearest distance to the center of our model structure. For instance if we set our lens at the redshift $z\simeq0.2$ then $t_2\simeq6.98 Gyr$ (in the flat matter dominated universe, the present time is about $9.18 Gyr$ if we set $H_0=71 {km}/{s Mpc}$). The density profile that we choose for this time is the universal halo density profile (NFW) \cite{NFW95} convolved with a negative Gaussian profile for mass compensation plus background density at that time:
\begin{equation}\label{ro2}
\rho(R,t_2) = \left(\rho_{NFW}-b_2 \rho_b(t_2)\right)e^{-\left(\frac{R}{R_2}\right)^2}+\rho_b(t_2),
\end{equation}
where
\begin{equation}
\rho_{NFW} = \rho_b(t_2)\frac{\delta_c}{\left(\frac{R}{R_s}\right)\left(1+\frac{R}{R_s}\right)^2}
\end{equation}
and
\begin{equation}
\delta_c = \frac{200}{3}\frac{c^3}{\ln(1+c)-\frac{c}{1+c}}.
\end{equation}
In our numerical calculation we will use typical values $R_s=0.5Mpc$ and $c=5$ for a cluster. Note that at the time $t_2$ a black hole singularity covered by an apparent horizon has already been evolved. Therefore, the NFW profile has to be modified and a black hole mass greater than a minimum value has to be added to it at the center. This physical fact is reflected in a shell crossing singularity if we take the familiar NFW profile similar to that assumed for the time $t_1$. The mass we have assumed for this black hole singularity is about one thousandth of the mass up to the $R_s$ and equal to $5.66\times10^{11} M_{\odot}$. Figs. \ref{emclnfw} and \ref{tnclnfw} shows the LTB functions $E$ and $t_n$ as a result of these boundary assumptions. Using these LTB functions, the density profile of our model structure is obtained as depicted in Fig. \ref{nfwden}.

\section{Gravitational Lensing in the dynamical setting}\label{GLLTB}

Our aim now is to study the path of a light ray emitting from a distant source far from the observer within our expanding inhomogeneous universe passing a gravitational lens. The result will then be compared with the familiar thin lens approximation in a FRW universe. This will show us how far this approximation is valid and where one has to expect deviations. To do this we have to explore suitable numerical methods to integrate the geodesic equations. It turns out this is a non-trivial integration and special cares have to be taken. To solve these equations we need four initial conditions, given the light-like condition (\ref{null}). The freedom in the choice of the affine parameter reduces our initial conditions to three. The general relativistic model being defined now leads to the geodesic equations to be integrated numerically.

\subsection{definitions}

\begin{figure}[ht]
\includegraphics[width = \columnwidth]{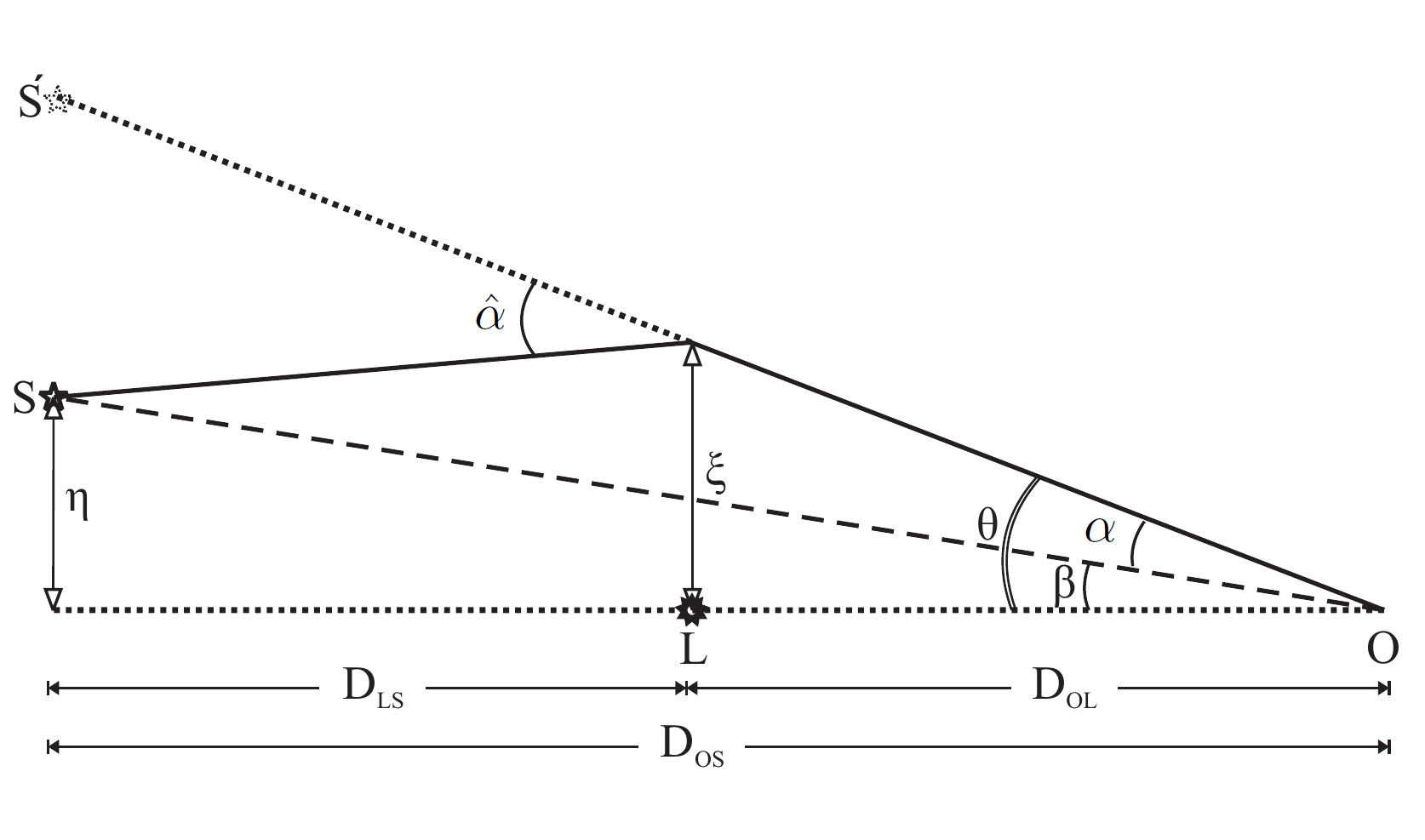}
\hspace*{10mm}\caption{\label{lensdiag}GL diagram: O is observer, S is source, S' is image in source plane, L is lens and $\hat{\alpha}$ is deflection angle.}
\end{figure}

The integration of the geodesics happens by a backshooting procedure. Our initial conditions are, therefore, the time of observation, distance of the observer from the lens expressed in terms of the redshift of the lens at the time of observation, and angle between the line of sight to the image of source and the line of sight to the lens or image angle at the observer location ($\theta$ in Fig. \ref{lensdiag}):\\
\begin{equation}
\tan\theta=\left.\frac{R\frac{d\phi}{d\lambda}}{R'\frac{dr}{d\lambda}}\right\vert_\text{null geodesic}.
\end{equation}
The integration is done from the observer to the source (at a specific redshift). Assuming then there is no lens, the model is a homogenous flat FRW universe and the geodesics are straight lines (in comoving coordinates) allowing us to determine angle between the source and the lens ($\beta$ in Fig. \ref{lensdiag}):
\begin{equation}
\tan\beta=\frac{\sin\phi_f}{\frac{r_i}{r_f}-\cos\phi_f},\\
\end{equation}
where $\phi_f$ is the $\widehat{OLS}$ angle, $r_i$ is the comoving distance of the observer, and $r_f$ is the comoving distance of the source from the center of coordinate system in the absence of the lens at the time $t_f$ defined by the root of the equation
\begin{equation}
\left(t_i^{\frac{1}{3}}-t_f^{\frac{1}{3}}\right)^2=\frac{1}{9}\left[\frac{R_i^2}{t_i^{\frac{4}{3}}}+ \frac{R_f^2}{t_f^{\frac{4}{3}}}-\frac{2R_i R_f}{t_i^{\frac{2}{3}}t_f^{\frac{2}{3}}}\cos\phi_f\right].
\end{equation}
We then may write the lens equation and determine the deflection angle $\hat{\alpha}$:
\begin{equation}
\hat{\alpha}=(\theta-\beta)\frac{D_{OS}}{D_{LS}},
\end{equation}
where we have assumed that the presence of the lens has not a significant effect on the distances and we may use the corresponding FRW distances. The reason that we calculate $\hat{\alpha}$ instead of $\alpha$ is that it is independent of distances and is a local quantity that depends only on the lens parameters.
\subsection{the numerical method}

\begin{figure}[ht]
\includegraphics[width = \columnwidth]{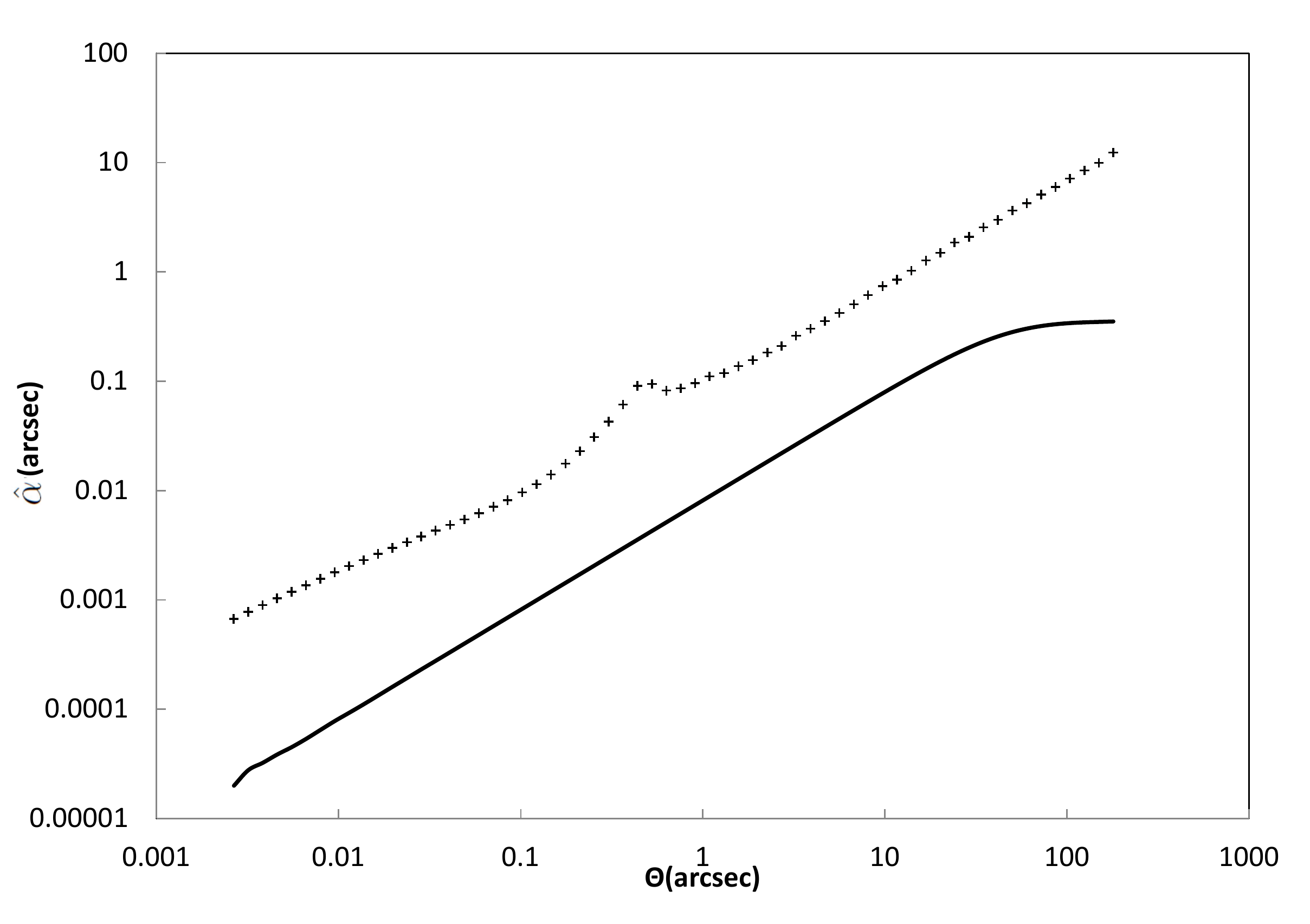}
\hspace*{10mm}\caption{\label{toyrk78}Deviation angle vs. image angle for diffuse density profile. Plus points are results of ray tracing by RK algorithm and solid line is for thin lens approximation.}
\end{figure}

\begin{figure}[ht]
\includegraphics[width = \columnwidth]{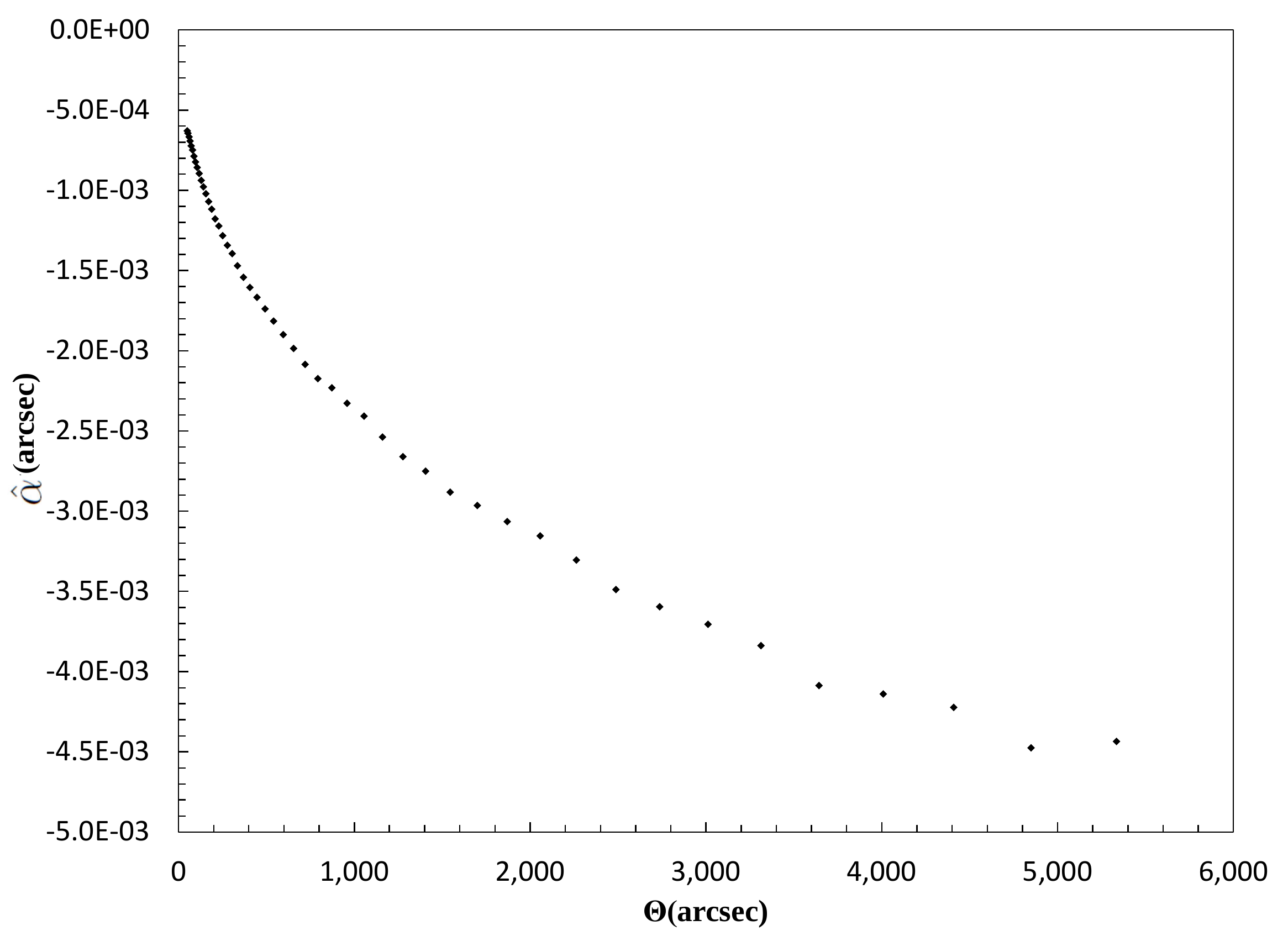}
\hspace*{10mm}\caption{\label{frwrk}Deviation angle in FRW model in Runge-Kutta integration method.}
\end{figure}

Given the complexity of the differential equations we intend to integrate, the numerical method we use may be very sensitive. We first started with the familiar Runge-Kutta adaptive step size algorithm with proportional and integral feedback (PI control) \cite{NR07} in which the step size is adjusted to keep local error under a suitable threshold. We started with the so-called embedded Runge-Kutta of the rank 5(4). It turned out, however, that its accuracy is too low. Therefore, we tried the rank 8(7) and then the rank 11(10) algorithm. The difference between these two last ranks, however, turned out to be marginal and below one percent, Given the time-consuming rank 11(10) algorithm, we preferred to use the rank 8(7) one. Now, as a first test for the accuracy of the numerical method we tried the trivial case of an LTB model namely the FRW case expecting a null result. A non-negligible deflection angle of the order of few miliarc seconds turned out as the result as shown in Fig. \ref{frwrk}. We suspected to face a numerical effect. To understand the numerical algorithm more deeply and the source of the effect more clearly, we continued to calculate a more concrete and non-trivial LTB case, and tried two different density profiles. The result for the rank 8(7) Runge-Kutta numerical method applied to a structure with a compact density profile did agree with the thin lens approximation. However, in the case of a more diffuse density profile the result showed a deflection angle up to an order of magnitude higher than the thin lens approximation, as shown in Fig. \ref{toyrk78}. Suspecting the effect to be not a genuine one but just a numerical artefact due to the errors in the LTB metric functions, we did increase the accuracy of that part of calculations related to the sensitive LTB functions. The problem, however, persisted even in the case of the toy model with analytic metric functions. We therefore did not trust the result and turned to other numerical methods! \\

\begin{figure}[ht]
\includegraphics[width = \columnwidth]{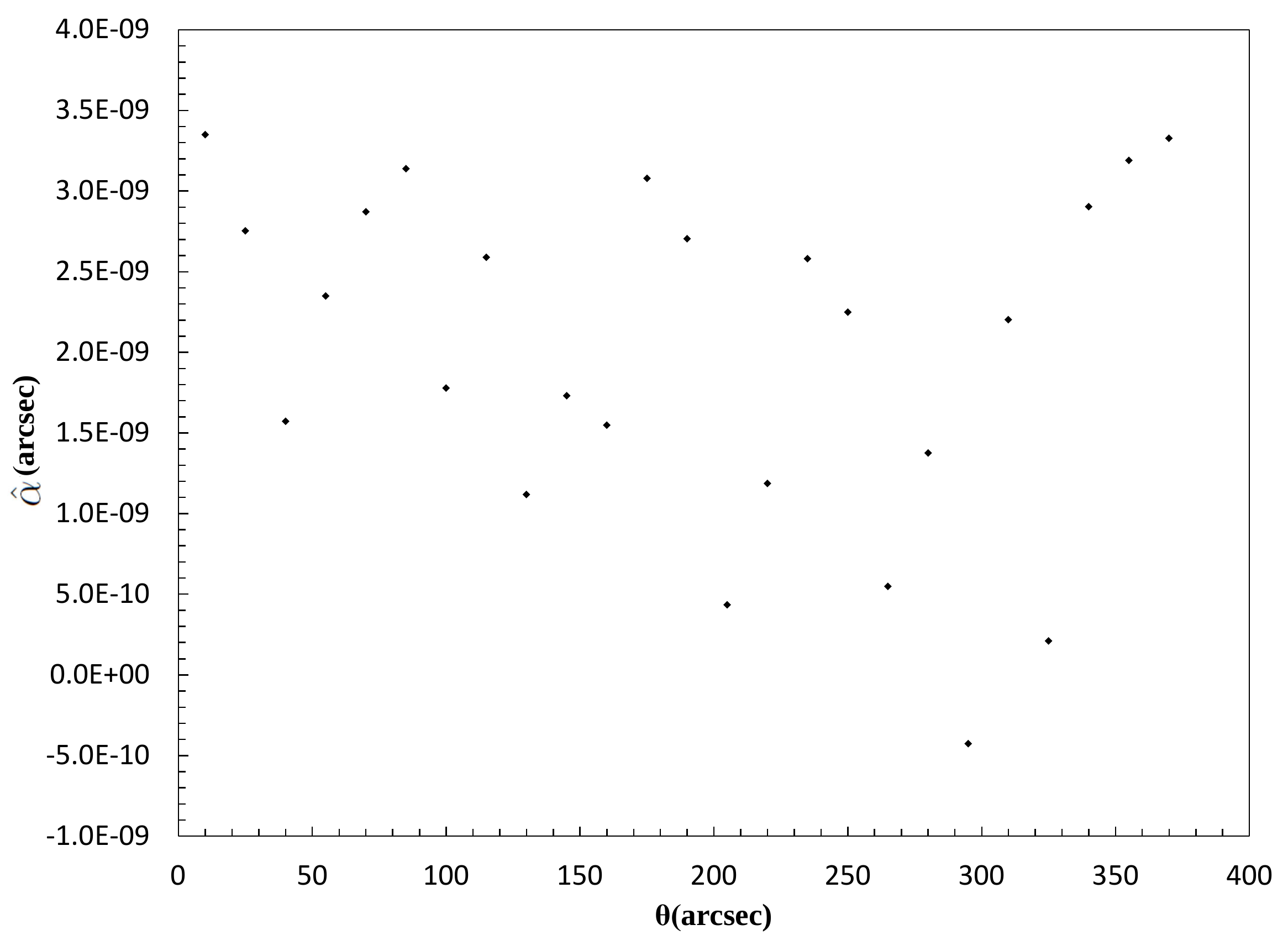}
\hspace*{10mm}\caption{\label{frwros}Deviation angle in FRW model in the Rosenbrock integration method, horizontal axis is the angle between image and lens.}
\end{figure}

The root of the numerical deficiency could be due to the term $\frac{d\phi}{d\lambda}$ in our equations, which is almost zero in the most part of the path of the light ray and changes suddenly to $\pi$ in the vicinity of the lens. This is a well-known phenomenon in the numerical calculation of the differential equations called as "stiff" \cite{DV84}. The characteristic property of such equations is the presence of two quite different scales. In our case we have on one side the cosmological distance scale of the source relative to the lens and observer and on the side the scale of the structure or the nearest distance of the ray to the lens. This led us to the so-called semi-implicit Rosenbrock method of the numerical integration of differential equations \cite{DV84,NR07}. As a first test we did again the trivial case of a FRW model. The result shown in Fig. \ref{frwros}, in contrast to Fig. \ref{frwrk}, is an acceptable one, showing how sensitive different numerical methods maybe. Therefore, we have done the integration of the geodesic equations by using the semi-implicit Rosenbrock method.

In the case of NFW profile, there are regions where the calculation of LTB metric functions needs more accuracy. Within the parametrization we are using, this occurs at the deep sector of the underdense region where the mass, say $r$, increases very slowly. We therefore decrease the mass step there. During the interpolation procedure we have checked if the results of the derivatives due to the interpolation coincides within the required accuracy with the independent results for the same derivatives according to numerical algorithm. In cases of large deviations we have increased the number of interpolating points to achieve the required accuracy.

\subsection{Thin Lens approximation}

Our aim is to compare the result of the exact model with the thin lens approximation to see if and how they my differ. The Fermat principle used in the thin lens approximation means that the light rays choose the minimum time. Within the gravitational lensing effect we expect, however, two different terms for the time delay due to different paths and due to different gravitational potentials. It is well known that in the presence of the gravitational field $\Phi$, clocks work slower by a factor of $(1+2\Phi)$. Given the assumption that the light path does not deviate from the background path except at the lens plane, the geometrical time delay can be approximated easily by the cosines law in the $OIS$ triangle of Fig. \ref{lensdiag}:

\begin{equation}\label{geotd}
    \Delta t_{geo}=\frac{D_{OL}D_{OS}}{2 D_{LS}}(\theta-\beta)^2
\end{equation}
To obtain the gravitational part, we first solve the geodesic equations of the thin lens approximation to first order in the gravitational field $\Phi$. The perturbed FRW metric in the Newtonian gauge is written as
\begin{equation}\label{FRWper}
    ds^2=-(1+2\Phi)dt^2+a^2(t)(1-2\Phi)(dr^2+r^2d\Omega^2).
\end{equation}

The corresponding gravitational time delay consists of two parts: a part which mainly depends on the metric and lens-source-observer configuration and is almost independent of the light trajectory or the image in the interested cases of small $\theta$. Therefore, it is usually canceled out while calculating the time delay for two images of the same configuration.  However, in our case, being interested in the time interval difference between a lensed image in an inhomogeneous LTB model and the corresponding unlensed image in a FRW model, one has to take this "configuration-term" into consideration\cite{weinberg2010}. Now, the time delay is given by
\begin{equation}
\Delta t_{grv} =\Psi(\theta)= \psi_I(\theta) + \psi_{II}(\theta),
\end{equation}
where
\begin{eqnarray}
\psi_I(\theta)&=&\int_{\xi}^{r_\infty}\frac{4\Phi(r')}{\sqrt{1-(\frac{\xi}{r'}})^2}dr'=2\int_{0}^{z_\infty}\Phi(\theta,z')dz', \nonumber \\
&=& \frac{1}{\pi}\int \kappa(\theta')\ln(|\theta-\theta'|)d\theta'^2 \label{defpot}.
\end{eqnarray}
Here $\kappa$ is the scaled surface mass density of the lens. In the case of axially symmetric case this term simplifies to
\begin{equation}\label{psiIax}
   \psi_{I}(\theta)=2\ln\theta \int_0^\theta \theta' \kappa(\theta')d\theta' + 2 \int_\theta^{\theta_\infty}\theta'\kappa(\theta')\ln\theta'd\theta'.
\end{equation}
Assuming a point like lens ($\Phi=-\frac{M}{r}$), this term for the gravitational time delay leads to $-4M\ln (\theta/2)$. If we assume, however, a NFW density profile, we obtain
\begin{equation}\label{defpotnfw}
\psi_{I,NFW}(x)=8\pi\rho_b\delta_c R_s^3\left(\ln^2\frac{x}{2}- \mathrm{arctanh}^2\sqrt{1-x^2}\right),
\end{equation}
where $x=R/R_s$. \\
The configuration-term of the time delay is given by
\begin{equation}\label{psiII}
\psi_{II}(\theta)=\int_\xi^{r_\infty}2(\Phi(r')-\Phi(\xi))\frac{\left(\frac{\xi}{r'}\right)^2}{\left(1+ \left(\frac{\xi}{r'}\right)^2\right)^{3/2}}dr'.
\end{equation}
It is easily seen that for a point like lens this term leads to $2M(1-\theta)$. \\

\begin{figure}[ht]
\includegraphics[width = \columnwidth]{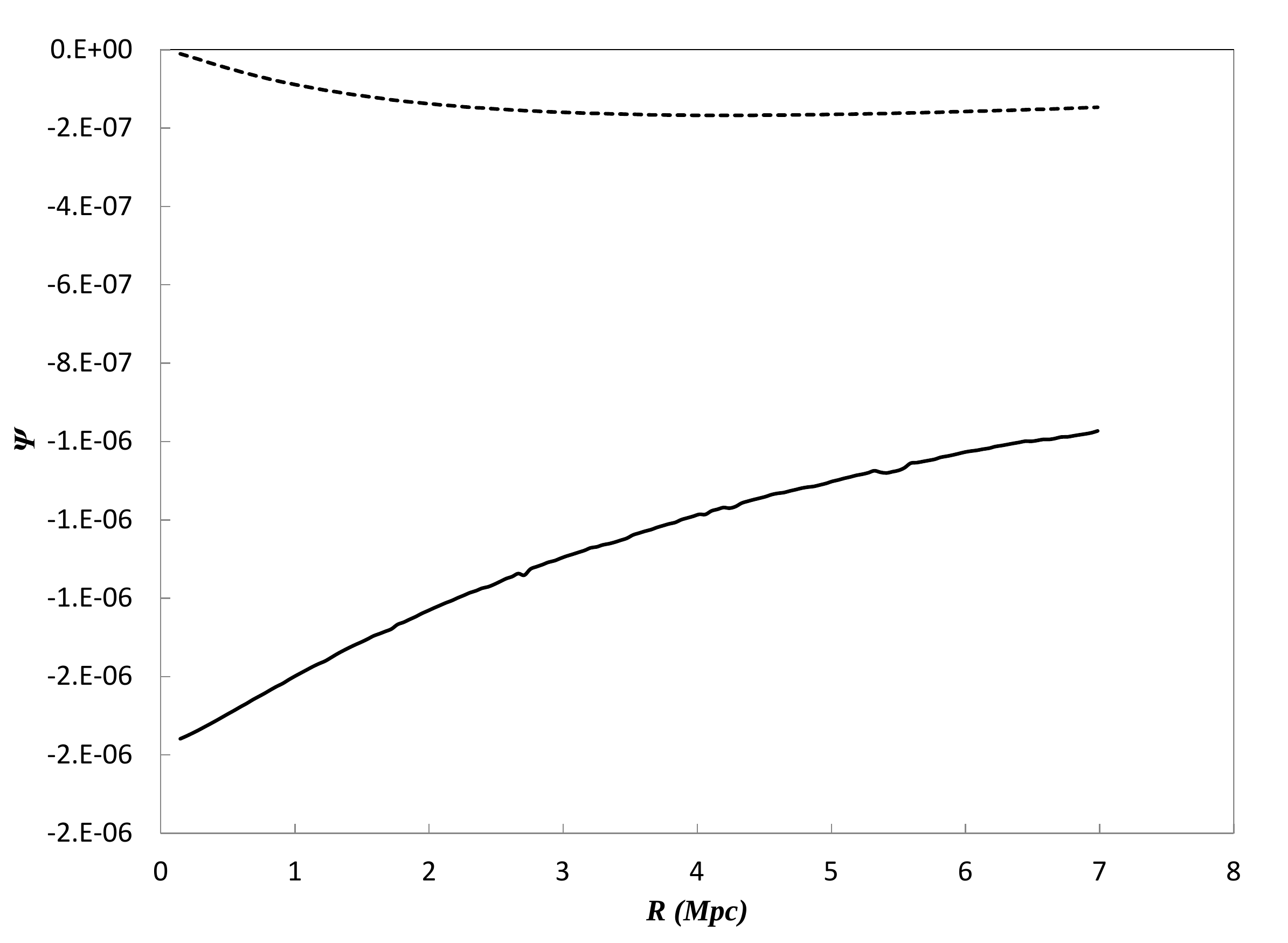}
\hspace*{10mm}\caption{\label{grtd}Gravitational time delays plotted against impact parameter $R$. Continues line is for part I and dashed lines depicted part II share.}
\end{figure}

We have calculated numerically the contribution of each of the above terms in the gravitational time delay using a NFW density profile. The result is depicted in Fig. \ref{grtd}. Evidently, the contribution of the second term is approximately constant. Therefore, in the usual case of measuring the difference of time delays for two or more images, this term may be ignored. However, given our interest of comparing time delays to the FRW case, we have to keep this term.\\

There is another issue we have to face. The two metrics (\ref{ltbm}) and (\ref{FRWper}) are not written in the same gauge. We have therefore to see how to transform the coordinates from the Newtonian gauge to synchronous LTB time and look for probable effects on the time delay we are interested in. This is done by K. Van Acoleyen \cite{acoleyen08} and A. Paranjape, T. P. Singh \cite{paranjape08}:
\begin{eqnarray}
t_{NFW}&=& t_{PFRW}+a(t_{PFRW})\int_r^\infty v_{pec}dr \label{ttrans} \\
v_{pec}&=&\dot{R}-\frac{\dot{a}}{a}R.
\end{eqnarray}
This transformation is related to the observer moving from infinity to the comoving distance $r$ and back again to infinity. It is therefore seen that the extra term in (\ref{ttrans}) is canceled out. Hence, the time delays in these two coordinates are equal, as expected due to the fact that both metrics tends to FRW metric at infinity. Now, neglecting the second term and applying the Fermat principle, we obtain the lens equation \cite{Virbhadra08}:
\begin{equation}\label{lenseqn}
\theta-\beta=\frac{D_{LS}}{D_{OL}D_{OS}}\frac{d\Psi(\theta)}{d\theta}
\end{equation}

\section{Results}\label{res}

Given the ambiguity of the mass definition in general relativity, the interpretation of the results require a more than usual details to be checked. That is why we have looked for different results corresponding to six different data sets, depending on the underlying model and assumptions to calculate the deflection angle, and on the mass associated to a definite density profile. We already know that there are different masses corresponding to one and the same density profile, none of them being preferred on the physical basis. \\

We first calculate different cases related to a toy model density profile. This will show us the characteristic features of different conditions of our model being based on the LTB or a thin lens- (TL-) model. We then go on to assume a NFW density profile and a structure of the scale of a galaxy structure and compare the results of LTB with TL model using different cutoffs. The time delay for different models and a comparison of its features relative to the deviation angle is also given.

\subsection{Deviation angle for the toy model}

\begin{figure}[ht]
\includegraphics[width = \columnwidth]{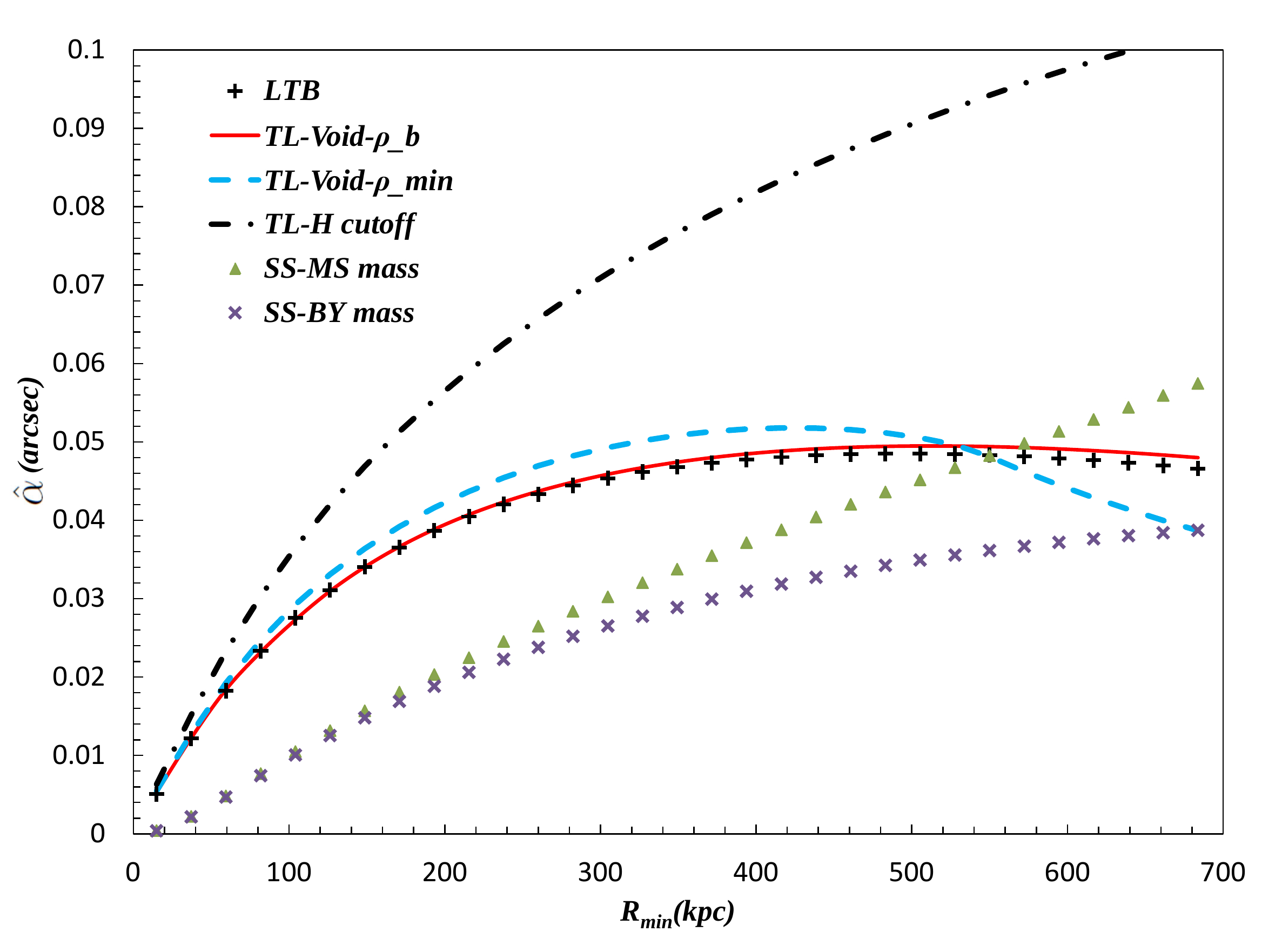}
\hspace*{10mm}\caption{\label{toydev} Deviation angles for the toy model. The thin lens approximation with a cutoff after the void is almost identical to the exact GR results. If we change the cutoff in the thins lens approximation to the point where the density is minimum at the  minimum of the void then there are some deviations from the our exact GR model. We may define the cutoff at the point where the expansion rate of the structure is equal to that of the background. Then the deviation from the exact GR result is increasing with the increase of teh $R_{min}$. Assuming a Schwarzschild point mass, the results depends heavily on the mass definition. The picture shows the result for the corresponding Misner-Sharp and Brown-York masses.}
\end{figure}

Fig. \ref{toydev} shows the results for the deflection angle as a function of the impact parameter $R_{min}$ expressed in kpc for different gravity model models and conditions but the same toy model for the density profile described in section \ref{structmodel}. In the following we describe theses cases differentiating them with the same symbol as in the figure.\\

\begin{description}
\item[$\boldsymbol{+}\;$] {\textbf LTB} \quad General relativistic exact model based on our LTB model solution.  The geodesic equations are integrated numerically. No other assumption is required. Note that at $R = 550 kpc$,  where almost all different curves coincide,  we have the void.
\item[\solidrule] {\textbf TL-Void-$\rho_b$} \quad Thin lens formalism based on a mass defined by the projection of the \emph{overdensity} $ \rho-\rho_b$ on the lens plane along the light path. Note that all the mass up to a distance where the density is almost equal to the background density after the void is included, i.e. including those points where the effective mass in negative due to the negative overdensity (underdensity).

\item[\protect\dashedrule] {\textbf TL-$\rho_{min}$} \quad Thin lens approximation based on a mass defined by projection of density on the lens plane and a cutoff at the minimum of the density ($\rho_{min}$), i.e. minimum of the void.
\item[\protect\dasheddotrule] {\textbf TL-$H$-cutoff} \quad Thin lens formalism and a cutoff defined by the location where expansion rate equals to background Hubble rate.
\item[$\boldsymbol{\blacktriangle}\;$] {\textbf SS-MS Mass} \quad Schwarzschild (point mass) model. The mass is chosen to be the Misner-Sharp mass.
\item[$\boldsymbol{\times}\;$] {\textbf SS-BY Mass}  \quad Schwarzschild model with the general relativistic Brown-York mass corresponding to the same density profile.
\end{description}

Before going to the more realistic NFW density profile we should look more into the details of the differences to the results  obtained for the toy model characteristic of the results for different cases relevant to astrophysical applications. The most relevant result is that the general relativistic exact model leads to almost identical result to the thin lens approximation  with the relevant density being the overdensity $\rho-\rho_b$ if one chooses the cut off at a point where the void matches to the  background. Otherwise, for example choosing the cutoff at the minimum density, we should expect some deviations. If the cut off is  chosen to be at the point where the expansion rate associated to the structure is equal to the the background Hubble parameter the deviation is much more and increases with the increase of the impact radius. We have added the result for a model based on the  mass concentrated at the center of the structure, the so called Schwarzschild case in the figure. We have chosen the relevant mass to  be the familiar Misner-Sharp or Brown-York one. The results are expectedly different and different to the  thin lens and LTB one. A  very interesting feature of these cases is that some of them coincide for the exact LTB model for the impact parameter being almost  equal to the  void distance.

\subsection{Deviation angle for the NFW density profile}

We now turn to a more realistic density profile, the NFW one, to make the same comparison between the exact LTB model and the thin lens approximation. We have first taken the density to be the oversdensity in an otherwise FRW model, namely $\rho-\rho_b$. We know already that the matching of any overdensity region to a background in general realtivity goes through an underdensity region or a void \cite{khakshournia}. In the case of exact LTB model this is implicitly done by the integration of the geodesics. In the NFW case this is usually dismissed by taking the NFW density to be the pure density and integrating it out to zero. We may, however, add a void to the NFW density profile and match it to the background as described in section \ref{NFW model}. In the case of the pure NFW without void, the corresponding equations can be integrated analytically to give the deviation angle \cite{bart96,keet02}:

\begin{eqnarray}
  \hat{\alpha}(x) &=& {\frac{4M_{sing}}{x R_s}}+{16\pi\rho_b \delta_c \frac{R_s^2}{x}}
  {\left( \log{\frac{x}{2}} + F(x)\right)}\label{nfwgamma}\\
  F(x)&=&\left\{\begin{array}{lr}
               \frac{\textrm{arctanh}({\sqrt{1-x^2}})}{\sqrt{1-x^2}}& x<1 \\
               1 & x=1 \\
               \frac{\arctan({\sqrt{x^2-1}})}{\sqrt{x^2-1}}& x>1
             \end{array}\right.
\end{eqnarray}

\begin{figure}[ht]
\includegraphics[width = \columnwidth]{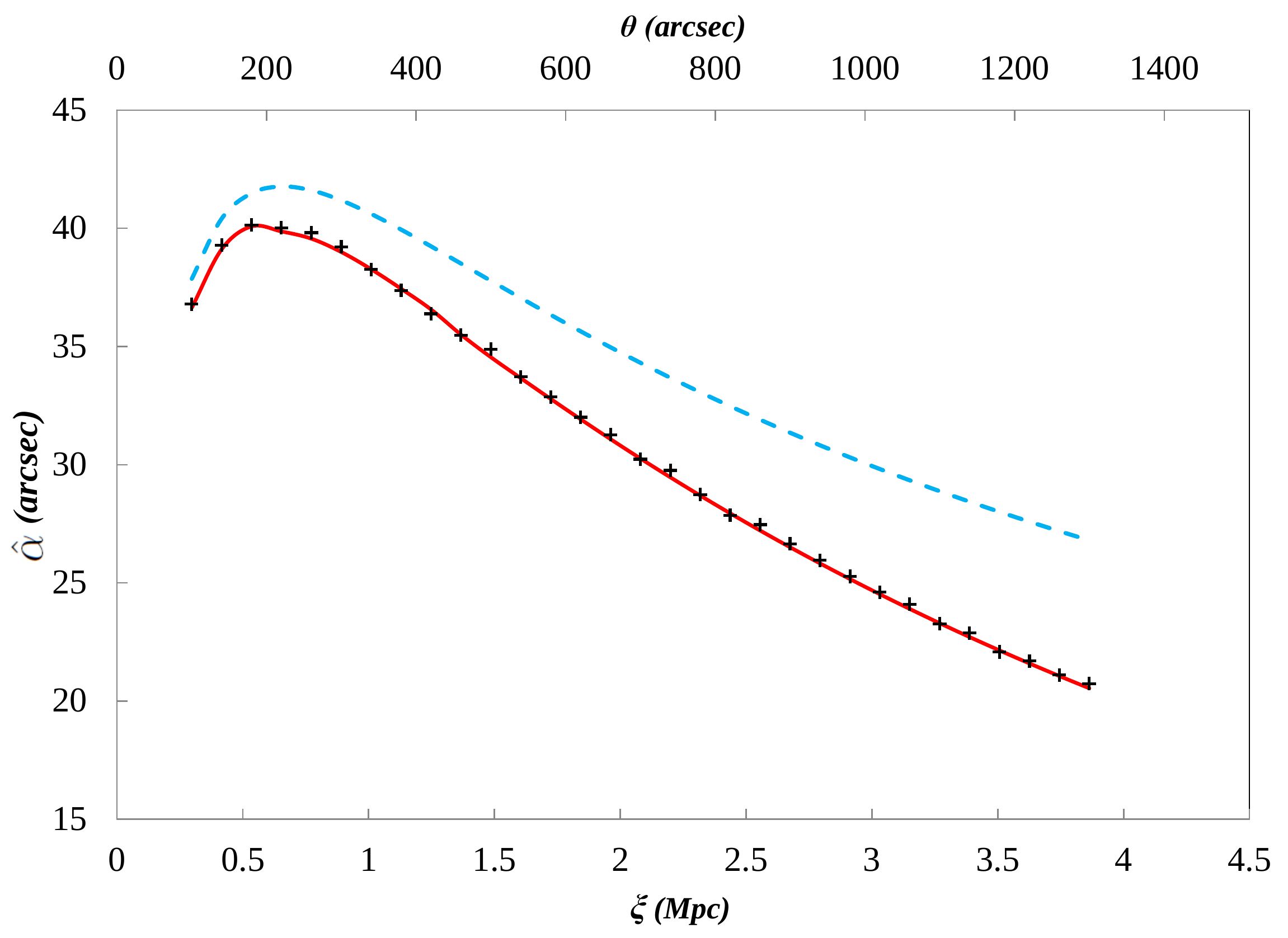}
\hspace*{10mm}\caption{\label{nfwdev}Deviation angle in NFW model. Plus points are from geodesic integration, continues line is for thin lens approximation and dashed line is for pure NFW profile without underdensity region (formula (\ref{nfwgamma})).}
\end{figure}

\begin{figure}[ht]
\includegraphics[width = \columnwidth]{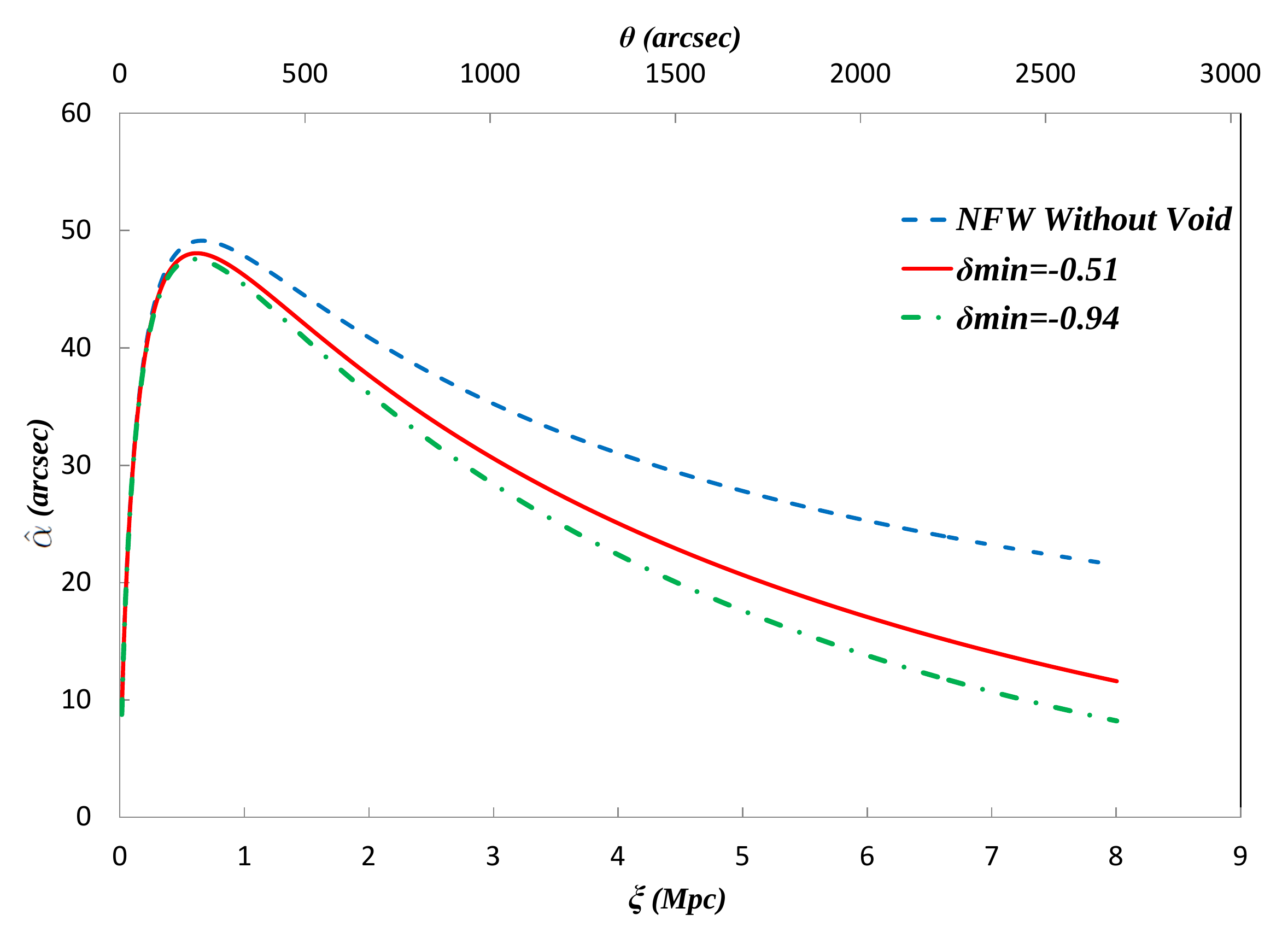}
\hspace*{10mm}\caption{\label{defdifvoid} Deviation angle for different voids. The dashed line is for the familiar NFW profile, the solid line for a void with the minimum density contrast of -0.51, and dashed dot line for a deeper void with the contrast -0.94 .}
\end{figure}

The result for the three cases, the exact LTB model with a modified NFW overdensity profile, thin lens approximation using the modified NFW with void, and the thin lens approximation using the normal NFW is depicted in Fig. \ref{nfwdev}. Here again the two cases of the thin lens approximation with NFW density profile including the void and the LTB exact method almost coincide. The thin lens approximation with the familiar density profile without a void differ however from the exact LTB model. The difference in the deviation angle can be more than 30 per cent depending on the impact parameter. The difference between the exact general relativistic LTB model and the thin lens approximation is due to the absence of the void in the normal NFW profile used in the thin lens approximation. To see the implications of the NFW parameters in this difference we have also calculated the deviation angle for different NFW profiles, with and without void. The result is depicted in the Fig. \ref{nfwgamma}. We see again that different NFW profiles including a void almost coincide with the exact LTB model. Models with the NFW profiles without void, however, differ substantially from the exact model. The difference is higher the bigger the $c$ parameter is, i.e. the less the concentration of the density of structure is. We are not aware of any observational evidence for such deviations. The effect does also depends on the specifications of the void profile, as shown in Fig. \ref{defdifvoid}. The deeper the void the more is the effect of deviation from the the original NFW profile. Note that the extension of the void is not independent of its depth due to the condition of mass compensation. The void density may also be changed by choosing different $b_2$ in \ref{ro2}.

\begin{figure}[ht]
\includegraphics[width = \columnwidth]{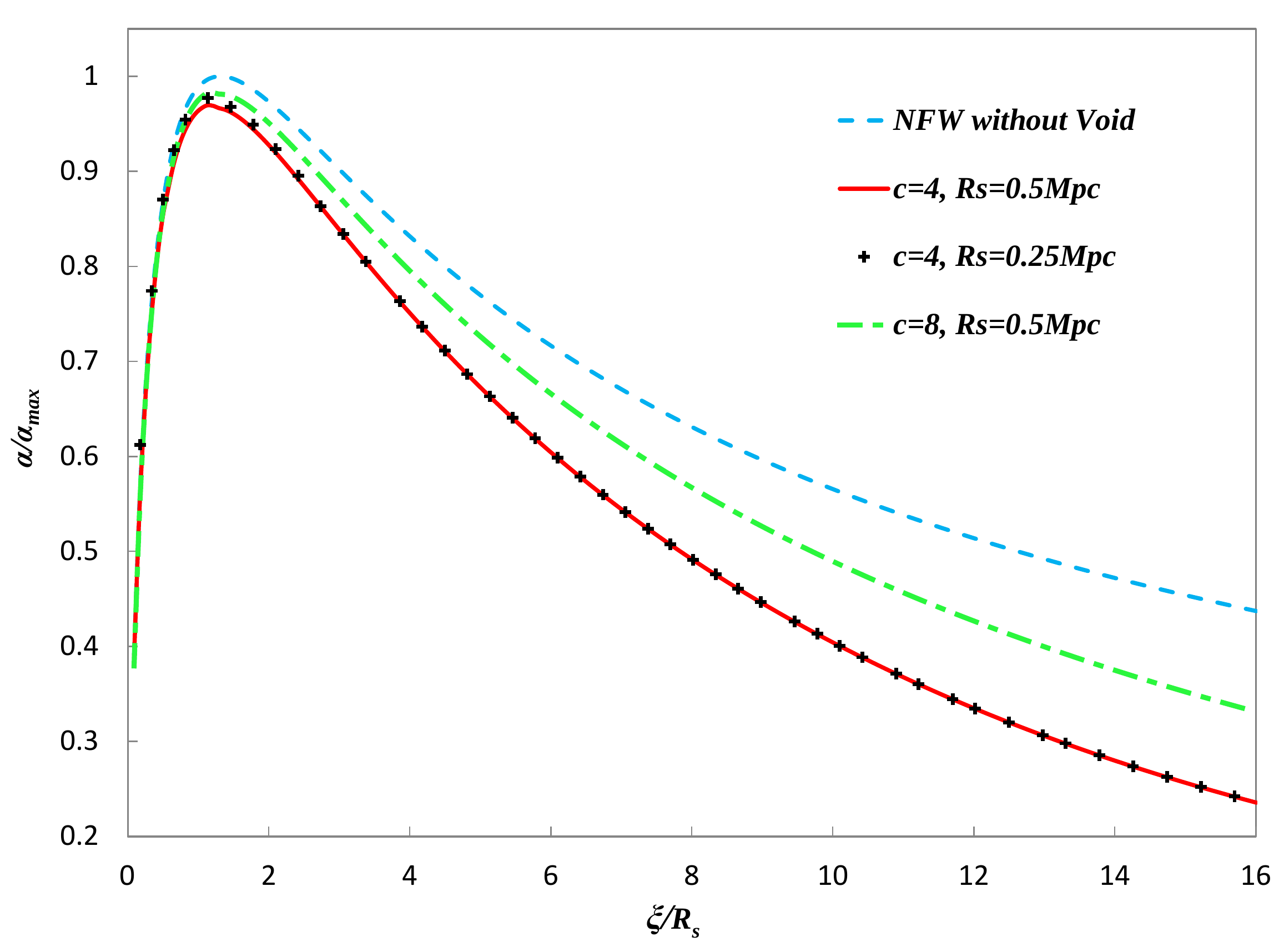}
\hspace*{10mm}\caption{\label{diffcrs}Deviation angle for the NFW model with different parameters. The horizontal axis is normalized to $R_s$ and the vertical axis to the maximum of the deflection angle for each model. The dashed line corresponds to the NFW model without void (formula (\ref{nfwgamma})). }
\end{figure}

The maximum deviation  of the deflection angle is therefore expected to occur at the large impact parameters.

\subsection{Strong lensing effects}

Tangential arcs as examples of observational objects produced by strong lensing may be used to study our results. It is, however, not a trivial task to use the observational data for our purposes. Usually the virial mass and concentration are reported. We had to convert them to $M_{200}$ and $c_{200}$ by the method mentioned in \cite{Johnston07}, assuming $\Omega_M=0.27, \Omega_\Lambda=1-\Omega_M$ and $h=0.7$). Table \ref{table} shows the effect of void on the modification of mass and concentration for some reported clusters. In all the cases the general relativistic effect is smaller than the errors. However, this modification increases with the increase of the Einstein radius. For example for A370, which has one of the farthest arc ever observed and has a low concentration, the effect of the void projection is the most.

\begin{table*}[t]
  \centering
  {\renewcommand{\arraystretch}{1.3}
 \begin{tabular}{ccccccccc}
 Cluster & $z_{cluster}$ & $z_{arc}$ & $M_{200}{\scriptstyle(10^{14}M_{\odot}h^{-1})}$ & $c_{200}$ & $\theta_{E}{\scriptstyle (arcsec)}$ & $M_{200}^{new}{\scriptstyle(10^{14}M_{\odot}h^{-1})}$ &  $c_{200}^{new}$ & Ref. \\
  \hline\hline
  SDSS1038+4849 & $0.430$ & $2.198$ & $0.70^{+0.49}_{-0.18}$ & $33.75^{+0.00}_{-18.46}$ & $12.6^{+1.3}_{-1.6}$ & $0.70$ & $33.79$ & \cite{oguri12} \\
  SDSS1209+2640 & $0.561$ & $1.021$ & $5.45^{+1.66}_{-1.31}$ & $6.69^{+1.38}_{-1.08}$ & $8.8_{-0.9}^{+0.9}$ & $5.49$ & $6.70$ & \cite{oguri12} \\
  A1703 & $0.277$ & $2.627$ & $9.46^{+1.66}_{-1.41}$ & $5.67^{+0.93}_{-0.69}$ & $27.4_{-2.7}^{+2.7}$ & $9.50$ & $5.68$ & \cite{oguri12} \\
  A370 & $0.375$ & $1.5$ & $2.58^{+0.32}_{-0.28}$ & $6.37^{+0.94}_{-0.77}$ & $43$& $2.62$ & $6.40$ & \cite{Broadhurst08} \\
  A1689 & $0.183$ & $3.05$ & $1.42^{+0.21}_{-0.20}$ & $12.40^{+3.23}_{-2.28}$ & $52$ & $1.43$ & $12.44$ & \cite{Broadhurst08} \\
  \hline
\end{tabular}}
\caption{\label{table}Modification of clusters mass and concentration by considering the effect of void on surface density based on arc observations ($\theta_E$ is Einstein radius)}
\end{table*}

 \subsection{Weak lensing effects}

The maximum deviation  of the deflection angle occurs at the large impact parameters. In practice, however, we expect weak lensing effects at such distances instead of the strong lensing, leading to modification of ellipticity of the background galaxies \cite{barwl}. The related observational quantity is the shear which in the axially symmetric cases is defined in as follows:
\begin{eqnarray}
  \gamma(\theta)=\frac{\overline{\Sigma}(\theta)-\Sigma(\theta)}{\Sigma_{crit}}\label{shear}\\
  \Sigma_{crit}=\frac{c^2}{4\pi G} \frac{D_{OS}}{D_{LS} D_{OL}}
\end{eqnarray}
For the familiar NFW profile we need only to calculate $\Sigma(\theta)$, due to the fact that the deviation angle is related to the average surface density by the $\alpha(\theta)= \theta \overline{\Sigma}(\theta)/\Sigma_{crit}$. For the surface mass density of the NFW profile ($x=R/R_s$) we have \cite{wright}:
\begin{equation}
\Sigma_{\rm nfw}(x) = \left\{ \begin{array}{ll}
\frac{2r_{s}\delta_{c}\rho_{c}}{\left(x^{2}-1\right)}
\left[1-\frac{2}{\sqrt{1-x^{2}}}{\rm arctanh}\sqrt{\frac{1-x}{1+x}}\hspace{0.15cm} \right]
 & \mbox{$\left(x < 1\right)$} \\
 & \\
\frac{2r_{s}\delta_{c}\rho_{c}}{3} & \mbox{$\left(x = 1\right)$} \\
 & \\
\frac{2r_{s}\delta_{c}\rho_{c}}{\left(x^{2}-1\right)}
\left[1-\frac{2}{\sqrt{x^{2}-1}}\arctan\sqrt{\frac{x-1}{1+x}}\hspace{0.15cm}
 \right]
& \mbox{$\left(x > 1\right)$}
\end{array}
\right.
\end{equation}

It is obvious from (\ref{shear}), that the shear is degenerate with respect to a constant surface mass density. Therefore, the statistical analysis of the ellipticity of the background galaxies is based on the reduced shear defined by \cite{barwl}:
\begin{equation}
g=\frac{\overline{\Sigma}(\theta)-\Sigma(\theta)}{\Sigma_{crit}-\Sigma(\theta)}\label{reducedshear}
\end{equation}

\begin{figure}[ht]
\includegraphics[width = \columnwidth]{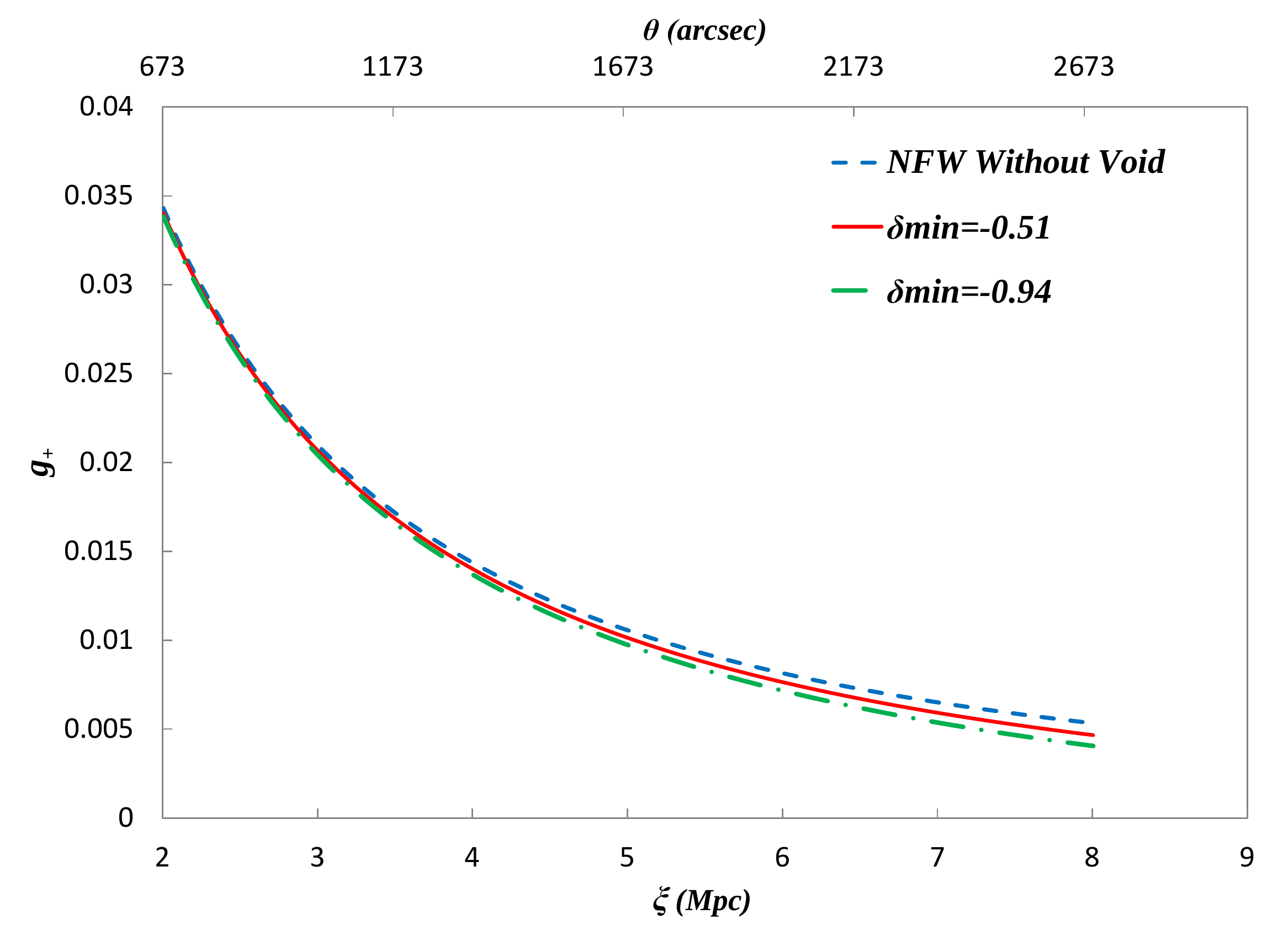}
\hspace*{10mm}\caption{\label{redshear}Reduced tangential shear for different profiles of the void. Dashed line is for the familiar NFW profile without void, the solid line corresponds to a void with the minimum density contrast of -0.51, and the dashed dot line for a deeper void with the contrast -0.94.}
\end{figure}

We have calculated the reduced shear for our modified NFW profile. The result shown in Fig.\ref{redshear} shows that there may be up to 10 percent difference at far distances.\\

\subsection{Time delay}

\begin{figure}[ht]
\includegraphics[width = \columnwidth]{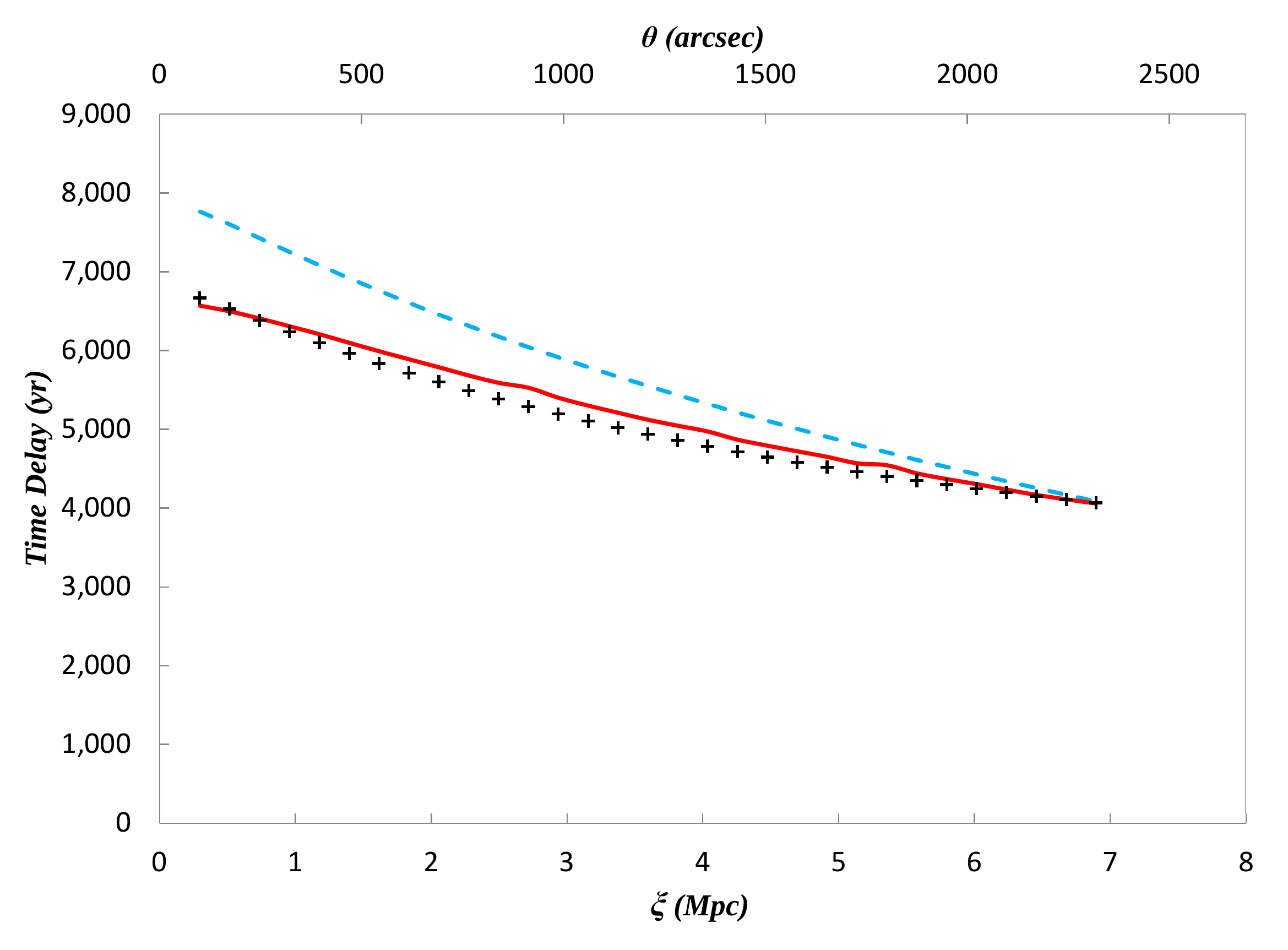}
\hspace*{10mm}\caption{\label{nfwtd}Time delay in NFW model. Plus points are for geodesic integration, continues line is for thin lens approximation and dashed line is for pure NFW profile (with the potential time delay from formula (\ref{defpotnfw})).}
\end{figure}

Fig. \ref{nfwtd} shows the time delay in three cases: exact LTB model and the thin lens approximation with the modified NFW density profile with a void, and the normal thin lens approximation with a NFW profile without a void. Here again we see that the geodesic integration method has a relatively good agreement to the thin lens approximation including a void, although the agreement is not as good as in the case of the deviation angle. In fact, the deviation angle is derived by differentiating the time delay relation. Therefore, it is not sensitive to an additive constant. In obsevational practice however, what is measured is the time delay between two or more images with the impact parameters of the same order.

\section{Conclusion} \label{concl}

In this paper we have studied the lensing effect on the deviation angle and time delay of cosmological light rays using an exact general relativistic LTB model for a structure within an FRW universe. The results are compared with the thin lens approximation used in the gravitational lensing literature for a toy model structure and a structure with a NFW density profile. We have also differentiated a normal NFW density profile going to zero and a modified one with a void before matching to the FRW background. The aim of the paper has been to study if and how viable the thin lens approximation based on the weak gravity in the lensing technology is. \\

The main results of the paper is reflected in the Figs. \ref{nfwdev}, \ref{nfwgamma}, and \ref{nfwtd}. We have shown that by using our modified NFW density profile the thin lens approximation coincides with the exact LTB general relativistic structure model to a high accuracy. Assuming a normal NFW density profile without a void, however, the thin lens approximation is not valid any more and the deviation angle may differ up to about 30 percent from the exact general relativistic calculation. This we have checked for different parameters of the NFW density profile. It turns out that the deviation increases with the increase of the $c$ parameter of the density profile, meaning that the deviation increases with the decrease of the compactness of the structure. Looking at some of the available observational data for strong and weak lensing the general relativistic effects seems to be within the accuracy of the observation.\\

We encounter a similar effect in the time delay of lensed images due to the gravitational lensing.\\

We conclude that in translating the weak gravity to different astrophysical phenomenon one has to take care of global relativistic features. We know from general relativity that, in contrast to the Newtonian dynamics, it is not possible to have an overdensity structure within an otherwise underdensity background except by the mediation of a void which compensates the overdensity \cite{khakshournia}. Now, the familiar technology of thin lensing using a NFW density profile dismiss this void which could easily be included in a modified density profile. Adding a void and using a modified NFW density profile correct this deficiency and one obtain a result almost identical to the exact general relativistic model. The deviation of the normal NFW density profile from the exact relativistic model depends on the compactness of the structure. The higher the compactness the less is the difference to the exact model. We should therefore see for such effects in the astrophysical observations of galaxy clusters of different compactness. This effect may influence our knowledge of the cluster masses and also the so-called concentration-mass relation of the galaxy clusters \cite{giocoli}. In the large survey operated by the Euclid mission several dozens of thousands of clusters up to the redshift $z=2$ is expected to be detectable. Therefore, we may be able not only to detect the effect we have calculated in this paper but also to use it to have a more precise estimate of the cosmological parameters.

\section{Acknowledgments}
We thank Dr. Sima Ghassemi for very useful discussions. Numerical calculations were carried out on Bahman cluster of the School of Astronomy, IPM, and the HPC lab of physics department of Sharif University of Technology.

\section{Appendix}\label{app}
For the case of $E(r)=0$ we have:
For the case of $E(r)=0$ we have:
\begin{flalign}
&R=\left(\frac{9}{2}M(t-t_{n})^2\right)^{\frac{1}{3}}\\
&\dot{R}=\frac{2}{3}\frac{R}{t-t_n}\\
&R'=\frac{1}{3}R\frac{M'}{M}-\frac{2}{3}R\left(\frac{t'_n}{t-t_n}\right)\\ \nonumber
\end{flalign}
\begin{flalign}
&\dot{R'}=\frac{2}{3}R\left(\frac{1}{t-t_n}\right)\left(R'+R\frac{t'_n}{t-t_n}\right)\\
&R''=\frac{1}{3}R'\frac{M'}{M} + \frac{1}{3}R \left(\frac{M''}{M}-\frac{M'^2}{M^2}\right) \nonumber\\
&- \frac{2}{3}R'\left(\frac{t'_n}{t-t_n}\right) - \frac{2}{3}R\left(\frac{t''_n}{t-t_n}+ \frac{t'_n}{(t-t_n)^2}\right)
\end{flalign}
and for $E(r)<0$:
\begin{flalign}
&R=-\frac{M}{E}(1-\cos\eta)\\
&\eta-\sin\eta=\frac{(-E)^{3/2}}{M}(t-t_{n}(r))\\
&\dot{R}= (-E)^\frac{1}{2}\frac{\sin\eta}{1-\cos\eta}\\
&R'=\dot{R}(t-t_n)\left(\frac{3}{2}\frac{E'}{E}-\frac{M'}{M}\right) + R\left(\frac{M'}{M}-\frac{E'}{E}\right) -\dot{R}t'_n\\
&\dot{R'}= \frac{M}{R^2}\left[t'_n - (t-t_n)\left(\frac{3}{2}\frac{E'}{E}-\frac{M'}{M}\right)\right] + \frac{1}{2}\frac{E'}{E}\dot{R}\\
&R''= R'\left(\frac{M'}{M}-\frac{E'}{E}\right)+ R\left[\frac{M''}{M}-\frac{M'^2}{M^2}-\left(\frac{E''}{E} -\frac{E'^2}{E^2}\right)\right] \nonumber\\
&+ \dot{R}\left[\frac{3}{2}\left(\frac{E''}{E}- \frac{E'^2}{E^2}\right)- \left(\frac{M''}{M}-\frac{M'^2}{M^2}\right)\right](t-t_n) \nonumber\\
&+ \left(\frac{3}{2}\frac{E'}{E}-\frac{M'}{M}\right)\left[\dot{R'}(t-t_n)-\dot{R}t'_n\right] - \dot{R'}t'_n - \dot{R}t''_n\\
&\frac{X'}{X}= \frac{R''}{R'}-\frac{E'}{2(1+E)}
\end{flalign}
and for $E(r)>0$ is similar to $E(r)<0$ only $\sin$ and $\cos$ are transformed to hyperbolic functions $\sinh$ and $\cosh$.


\begin{thebibliography}{}
\bibitem{GLenses}
P. Schneider, J. Ehlers, E.E. Falco, \emph{Gravitational Lenses}, Springer-Verlag (1992).
\bibitem{Sas93}
M. Sasaki, Prog. Theor. Phys., \textbf{90}, No. 4 (1993).
\bibitem{Fut95}
T. Futamase, Prog. Theor. Phys., \textbf{93}, No. 3 (1995).
\bibitem{FriKling11}
S. Frittelli, T. P. Kling, Mon. Not. R. Astron. Soc., \textbf{415}, 3599-3608 (2011).
\bibitem{DR73}
C. C. Dyer, R. C. Roeder, Astrophysical Journal, \textbf{180}, L31, (1973).
\bibitem{Marra:2011ct}
V.~Marra and A.~Notari, Class.\ Quant.\ Grav.\  {\bf 28}, 164004 (2011).
\bibitem{GreenWald}
S. R. Green, R. M. Wald, Phys. Rev. D, \textbf{85}, 063512 (2012).
\bibitem{Szabados}
L. B. Szabados, Living Rev. Relativity, \textbf{4}, (2004).
\bibitem{taghizadeh}
J. T. Firouzjaee, M. Parsi Mood, R. Mansouri, Gen. Rel. Grav., \textbf{44}, 639 (2012).
\bibitem{khakshournia}
S. Khakshournia, R. Mansouri, Phys. Rev. D, \textbf{65}, 027302, (2001).
\bibitem{Bolejko12}
K.~Bolejko, C.~Clarkson, R.~Maartens, D.~Bacon, N.~Meures and E.~Beynon, Phys.\ Rev.\ Lett.\  {\bf 110}, 021302 (2013).
\bibitem{Lem97}
G. A. Lema\^{\i}tre, Gen. Rel. Grav., \textbf{29}, 5 (1997)(reprint).
\bibitem{Tol34}
R. C. Tolman, Proc. Nat. Acad. Sci., \textbf{20}, 169 (1934).
\bibitem{Bon47}
H. Bondi, Mon. Not. R. Astron. Soc., \textbf{107}, 410 (1947).
\bibitem{KH01}
A. Krasi\'{n}ski, C. Hellaby, Phys. Rev. D, \textbf{65},023501 (2001).
\bibitem{BKCH}
K. Bolejko, A. Krasi\'{n}ski, M. C\'{e}l\'{e}rier, C. Hellaby, \emph{Structures in the Universe by Exact Methods: Formation, Evolution, Interactions}, Cambridge University Press (2010).
\bibitem{NFW95}
J. F. Navarro, C. S. Frenk, S. D. M. White, Mon. Not. R. Astron. Soc., \textbf{275}, 720 (1995).
\bibitem{NR07}
W. H. Press, S. A. Teukolsky, W. T. Vetterling, B. P. Flannery, \emph{Numerical Recipes: The Art of Scientific Computing}, 3rd Edition, Cambridge University Press (2007).
\bibitem{DV84}
K. Dekker, J. G. Verwer, \emph{Stability of Runge-Kutta methods for stiff nonlinear differential equations}, North-Holland (1984).
\bibitem{weinberg2010}
S. Weinberg, \emph{Cosmology}, Oxford University Press, (2008).
\bibitem{acoleyen08}
K. Van Acoleyen, J. Cosmol. Astropart. Phys., \textbf{10}, 028, (2008).
\bibitem{paranjape08}
A. Paranjape, T. P. Singh, J. Cosmol. Astropart. Phys., \textbf{03}, 023, (2008).
\bibitem{Virbhadra08}
K.~S.~Virbhadra, Phys.\ Rev.\ D {\bf 79}, 083004 (2009).
\bibitem{bart96}
M. Bartelmann, Astron. Astrophys., \textbf{313},697 (1996).
\bibitem{keet02}
C. R. Keeton, arXiv:astro-ph/0102341
\bibitem{barwl}
M. Bartelmann, P. Schneider, Phys. Rep., \textbf{340}, 291, (2001).
\bibitem{wright}
C. O. Wright, T. G. Brainerd, Astrophys. J., \textbf{534}, 34, (2000).
\bibitem{oguri12}
M. Oguri, M. B. Bayliss, H. Dahle, K. Sharon, M. D. Gladders, P. Natarajan, J. F. Hennawi, B. P. Koester, Mon. Not. R. Astron. Soc., \textbf{420}, 3213-3239 (2012).
\bibitem{Broadhurst08}
T.~Broadhurst, K.~Umetsu, E.~Medezinski, M.~Oguri and Y.~Rephaeli, Astrophys. J.  \textbf{685}, L9 (2008).
\bibitem{Johnston07}
D.~E.~Johnston {\it et al.}, [SDSS Collaboration],arXiv:0709.1159 [astro-ph].
\bibitem{giocoli}
C. Giocoli, M. Meneghetti,S. Ettori,L. Moscardini, Mon. Not. R. Astron. Soc., \textbf{426}, 1558, (2011).
\end{thebibliography}

\end{document}